\documentclass[pdftex,twocolumn,epjc3]{svjour3}  

\usepackage{textcomp}
\usepackage{siunitx}

\usepackage[T1]{fontenc} 
\makeatletter
\renewcommand{\subsection}{%
  \@startsection{subsection}
    {2}
    {\z@}
    {-21dd plus-8pt minus-4pt}
    {10.5dd}
    {\normalsize\bfseries\boldmath}%
}
\makeatother
\smartqed  

\usepackage{graphicx} 
\usepackage{subcaption}
\usepackage[numbers,sort&compress]{natbib} 
\usepackage{hyperref} 
\hypersetup{colorlinks,citecolor=blue,urlcolor=blue,linkcolor=blue}
\usepackage{amssymb}
\usepackage[switch]{lineno}
\usepackage[usenames,dvipsnames]{xcolor}
\usepackage[normalem]{ulem}
\usepackage{siunitx}

\usepackage{blindtext}

\newcommand{\unitRad}{\SI[per-mode = symbol]{}{\,\micro\becquerel/kg}}
\newcommand{\mum}{\SI[per-mode = symbol]{}{\,\micro\meter}}
\newcommand{\mus}{\SI[per-mode = symbol]{}{\,\micro\second}}
\newcommand{\mmu}{\SI[per-mode = symbol]{}{\,\micro}}
\begin{document}
\title{\vspace{-2mm}The XENONnT Dark Matter Experiment}
\journalname{Eur.~Phys.~J.~C}
\author{E.\,Aprile\thanksref{add::columbia} \and
J.\,Aalbers\thanksref{add::nikhef} \and
K.\,Abe\thanksref{add::tokyo} \and 	
S.\,Ahmed Maouloud\thanksref{add::paris} \and			
L.\,Althueser\thanksref{add::munster} \and
B.\,Andrieu\thanksref{add::paris} \and				
E.\,Angelino\thanksref{add::torino} \and
J.R.\,Angevaare\thanksref{add::nikhef} \and		
V.C.\,Antochi\thanksref{add::stockholm} \and	
D.\,Ant\'on Martin\thanksref{add::chicago}
% ...
(XENON Collaboration\thanksref{e4})
}

\authorrunning{XENON Collaboration}
\thankstext{deceased}{deceased}
\thankstext{alsoatcoimbrapoli}{also at: Coimbra Polytechnic - ISEC, 3030-199 Coimbra, Portugal}
\thankstext{alsoatuniheidelberg}{also at: Physikalisches Institut, Universit\"at Heidelberg, Heidelberg, Germany}
\thankstext{alsoatroma}{also at: INFN-Roma Tre, 00146 Roma, Italy}
\thankstext{alsoatusc}{also at: Information Sciences Institute, University of Southern California, Marina del Rey, CA 90292, USA}
\thankstext{alsoatwestlake}{now at: Department of Physics, School of Science, Westlake University, Hangzhou 310030, China}
\thankstext{alsoathongkong}{now at: School of Science and Engineering, The Chinese University of Hong Kong, Shenzhen 518172, China.}
\thankstext{e1}{\tt colijn@nikhef.nl}
\thankstext{e2}{\tt lgrandi@uchicago.edu}
\thankstext{e3}{\tt marc.schumann@physik.uni-freiburg.de}
\thankstext{e4}{\tt xenon@lngs.infn.it}

\institute{Physics Department, Columbia University, New York, NY 10027, USA\label{add::columbia} 
\and
Nikhef and the University of Amsterdam, Science Park, 1098XG Amsterdam, Netherlands\label{add::nikhef} 
\and
Kamioka Observatory, Institute for Cosmic Ray Research, and Kavli Institute for the Physics and Mathematics of the Universe (WPI), University of Tokyo, Higashi-Mozumi, Kamioka, Hida, Gifu 506-1205, Japan\label{add::tokyo} 
\and
LPNHE, Sorbonne Universit\'{e}, CNRS/IN2P3, 75005 Paris, France\label{add::paris} 
\and
Institut f\"ur Kernphysik, Westf\"alische Wilhelms-Universit\"at M\"unster, 48149 M\"unster, Germany\label{add::munster} 
\and
INAF-Astrophysical Observatory of Torino, Department of Physics, University of Torino and INFN-Torino, 10125 Torino, Italy\label{add::torino} 
\and
Oskar Klein Centre, Department of Physics, Stockholm University, AlbaNova, Stockholm SE-10691, Sweden\label{add::stockholm} 
\and
Department of Physics \& Kavli Institute for Cosmological Physics, University of Chicago, Chicago, IL 60637, USA\label{add::chicago} 
\and
New York University Abu Dhabi - Center for Astro, Particle and Planetary Physics, Abu Dhabi, United Arab Emirates\label{add::nyuad} 
\and  
INFN-Laboratori Nazionali del Gran Sasso and Gran Sasso Science Institute, 67100 L'Aquila, Italy\label{add::lngs} 
\and
Physik-Institut, University of Z\"urich, 8057  Z\"urich, Switzerland\label{add::zurich} 
\and
Department of Physics and Astronomy, Purdue University, West Lafayette, IN 47907, USA\label{add::purdue} 
\and
SUBATECH, IMT Atlantique, CNRS/IN2P3, Universit\'e de Nantes, Nantes 44307, France\label{add::subatech}   
\and
Department of Physics and Astronomy, University of Bologna and INFN-Bologna, 40126 Bologna, Italy\label{add::bologna} 
\and
Max-Planck-Institut f\"ur Kernphysik, 69117 Heidelberg, Germany\label{add::heidelberg} 
\and
Physikalisches Institut, Universit\"at Freiburg, 79104 Freiburg, Germany\label{add::freiburg} 
\and
Department of Particle Physics and Astrophysics, Weizmann Institute of Science, Rehovot 7610001, Israel\label{add::wis} 
\and
Department of Physics \& Center for High Energy Physics, Tsinghua University, Beijing 100084, China\label{add::tsinghua} 
\and
LIBPhys, Department of Physics, University of Coimbra, 3004-516 Coimbra, Portugal\label{add::coimbra} 
\and
Department of Physics ``Ettore Pancini'', University of Napoli and INFN-Napoli, 80126 Napoli, Italy\label{add::naples} 
\and
IJCLab, CNRS/IN2P3, Universit\'e Paris-Saclay, Universit \'e de Paris, 15 rue Georges Cl\'emenceau, 91400, Orsay, France\label{add::ijclab} 
\and
Institute for Astroparticle Physics, Karlsruhe Institute of Technology, 76021 Karlsruhe, Germany\label{add::kit} 
\and 
Department of Physics and Astronomy, Rice University, Houston, TX 77005, USA\label{add::rice} 
\and
Department of Physics and Chemistry, University of L'Aquila, 67100 L'Aquila, Italy\label{add::laquila} 
\and
Institut f\"ur Physik \& Exzellenzcluster PRISMA$^{+}$, Johannes Gutenberg-Universit\"at Mainz, 55099 Mainz, Germany\label{add::mainz} 
\and
Department of Physics, University of California San Diego, La Jolla, CA 92093, USA\label{add::ucsd} 
\and
Kobayashi-Maskawa Institute for the Origin of Particles and the Universe, and Institute for Space-Earth Environmental Research, Nagoya University, Furo-cho, Chikusa-ku, Nagoya, Aichi 464-8602, Japan\label{add::nagoya} 
\and
Department of Physics, Kobe University, Kobe, Hyogo 657-8501, Japan\label{add::kobe} 
\and
INFN-Ferrara and Dip.~di Fisica e Scienze della Terra, Universit\`a di Ferrara, 44122 Ferrara, Italy\label{add::ferrara} 
}

\date{} 

\maketitle


\sloppy

\begin{abstract}
The multi-staged  XENON program at INFN Laboratori Nazionali del Gran Sasso aims to detect dark matter with two-phase liquid xenon time projection chambers of increasing size and sensitivity. The XENONnT experiment is the latest detector in the program, planned to be an upgrade of its predecessor XENON1T. It features an active target of 5.9\,tonnes of cryogenic liquid xenon (8.5\,tonnes total mass in cryostat). The experiment is expected to extend the sensitivity to WIMP dark matter by more than an order of magnitude compared to XENON1T, thanks to the larger active mass and the significantly reduced background, improved by novel systems such as a radon removal plant and a neutron veto. This article describes the XENONnT experiment and its sub-systems in detail and reports on the detector performance during the first science run.
\end{abstract}

\section{Introduction}
\label{sec::introduction}

XENON is a multi-staged research program that aims to detect dark matter using two-phase liquid xenon time projection chambers (LXe TPCs) of increasing size and sensitivity~\cite{Aprile:2010bt,Aprile:2011dd,Aprile:2017aty}. While these detectors are designed for high-sensitivity direct searches for dark matter in the form of Weakly Interacting Massive Particles (WIMPs), they also allow for carrying out a rich and diverse science program~\cite{Aprile:2018dbl,PhysRevLett.122.071301,PhysRevLett.122.141301,Aprile:2019jmx,Aprile:2019xxb,XENON:2019dti,PhysRevD.102.072004}. With an active LXe target mass of 5.9\,tonnes, the XENONnT experiment is the latest and largest detector of the program, designed to improve upon the sensitivity of XENON1T by more than an order of magnitude~\cite{XENON:2020kmp}.

To reach this sensitivity, XENONnT features a significantly reduced background level in a central 4\,tonnes fiducial volume compared to that observed in XENON1T. The LXe-intrinsic electron recoil background, dominated by $^{222}$Rn and $^{85}$Kr, was successfully reduced by about an order of magnitude~\cite{PhysRevLett.129.161805}. Material selection and a new neutron veto suppress the neutron-induced background~\cite{XENON:2023cxc}. 
The background requirements guided the XENONnT design as well as the construction and preparation procedures presented here.

\begin{figure*}[t]
\centering 
\includegraphics[width=.95\textwidth]{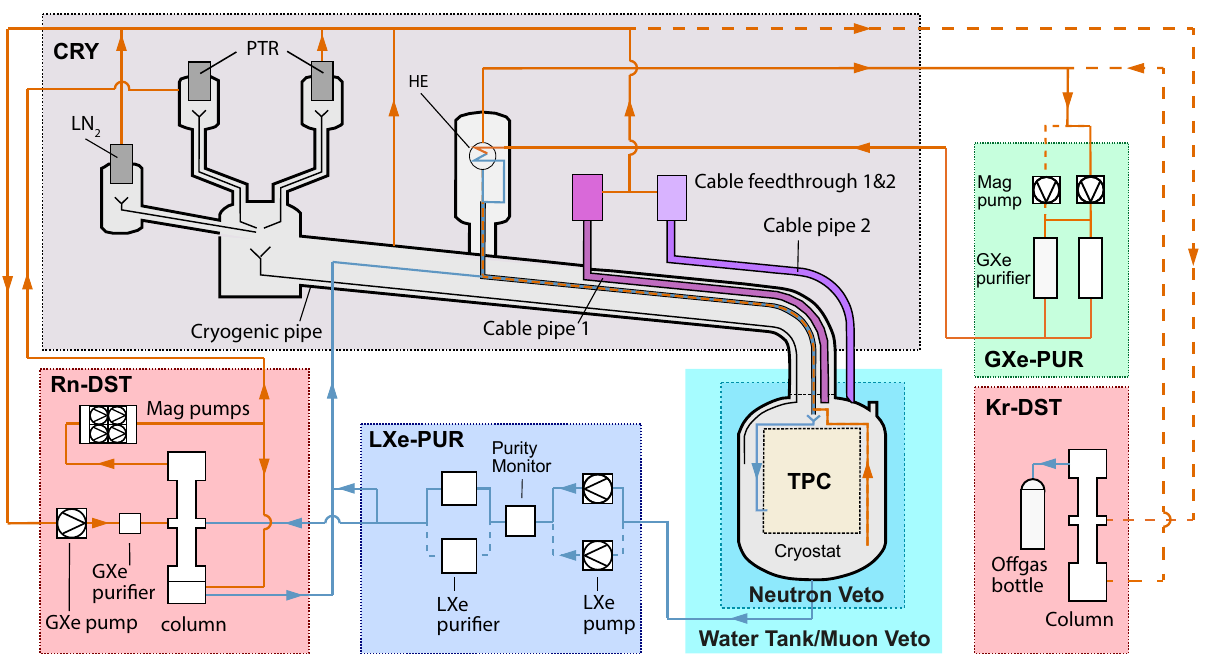}
\hfill
\caption{\label{fig:globalscheme}Schematic of the XENONnT experiment highlighting the detectors (TPC, neutron veto, muon veto) as well as the sub-systems for cryogenics and purification. Orange and blue arrows are used to show the path of gaseous and liquid xenon, respectively. The solid lines indicate the default flow path during normal operation, while the dashed lines indicate redundant flow paths and the optional flow path to allow for online krypton distillation. Details on the cryogenic system (CRY), the gaseous and liquid purification systems (GXe-PUR and LXe-PUR), and the krypton and radon removal plants (Kr-DST and Rn-DST) are reported in Section~\ref{sec::xehandling}. The TPC and the veto systems are described in Sec.~\ref{sec::tpc} and Sec.~\ref{sec::vetoes}, respectively.}
\end{figure*}

\section{The XENONnT Experiment}
\label{sec::xent}

A schematic of XENONnT highlighting its major sub-systems is shown in Fig.~\ref{fig:globalscheme}. The experiment is located in Hall~B of the INFN Laboratori Nazionali del Gran Sasso (LNGS) in Italy. It consists of three nested detectors next to a three-story service building containing the auxiliary systems needed to operate the detectors. XENONnT was planned as a fast upgrade of XENON1T and thus inherited much of the infrastructure from its predecessor. At the same time, it was equipped with several innovative features to improve the detector performance, lower the backgrounds, and extend its physics reach. These features are listed here.
    
\paragraph{LXe TPC:} The TPC accommodates a 5.9\,tonnes active LXe target. Particular emphasis went into (1) minimizing the use of materials (mostly Polytetrafluoroethylene - PTFE), (2) minimizing the number of individual components, (3) developing a novel field-shaping system to improve the uniformity of the electric drift field while minimizing charge-up effects and (4) characterizing all photomultiplier tubes (PMTs) in LXe to mitigate the risk of failure. See Sec.~\ref{sec::tpc} for more details.

\paragraph{LXe recirculation/purification:} XENONnT uses a novel liquid xenon purification system that complements the gas purification loop remaining from XENON1T. This greatly improves removing electronegative impurities, which was found the be challenging in XENON1T.  The system extracts LXe from the bottom of the cryostat with cryogenic liquid pumps at a rate between 4-16\,tonne/day and purifies it through custom-developed cryogenic, low-radon-emanation purifiers to remove electronegative impurities. See Sec.~\ref{sec::LXe_pur} for more details.

\paragraph{High-flow radon removal plant:} 
$^{222}$Rn is continuously emanated by all materials inside the detector systems. Decays of its daughter $^{214}$Pb constitute the dominant background source in tonne-scale LXe detectors. 
To reduce the radon-induced concentration to the design goal of $\sim$1\,\unitRad{}, XENONnT is equipped with a high-flow radon distillation plant that is operated continuously at a rate of  1.73~tonnes/day~\cite{Murra:2022mlr}. The liquid recirculation system extracts LXe directly from the cryostat and transfers it to the plant which efficiently removes radon. The plant can simultaneously process a flow of gaseous xenon to efficiently clean it when extracted from locations with high radon emanation. See Sec.~\ref{sec::rn} for more details.
 
\paragraph{Neutron Veto:} To reduce the neutron-induced nuclear recoil background, XENONnT features a veto system to tag neutrons. It encompasses a water volume of about 33\,m$^3$ around the cryostat that is optically separated from the water Cherenkov muon veto detector by reflective foils, but uses the same water.
Thermalized neutrons are subsequently captured under the emission of $\mathcal{O}$(MeV) $\gamma$-rays.  
These gammas accelerate electrons to relativistic speeds by Compton-scattering, yielding Cherenkov light for delayed tagging. 
After an initial run with demineralized water to characterize the TPC performance in a low-background environment, at the end of 2023 the water was loaded with gadolinium sulphate octahydrate to increase the neutron capture efficiency. See Sec.~\ref{sec::nveto} for more details.

\paragraph{}In the following, all XENONnT sub-systems will be described in detail, emphasizing those that are new or that underwent a major upgrade since XENON1T~\cite{Aprile:2017aty}. 

\subsection{The two-phase LXe TPC}
\label{sec::tpc}

\begin{figure*}[t]
\centering 
\includegraphics[width=.98\textwidth]{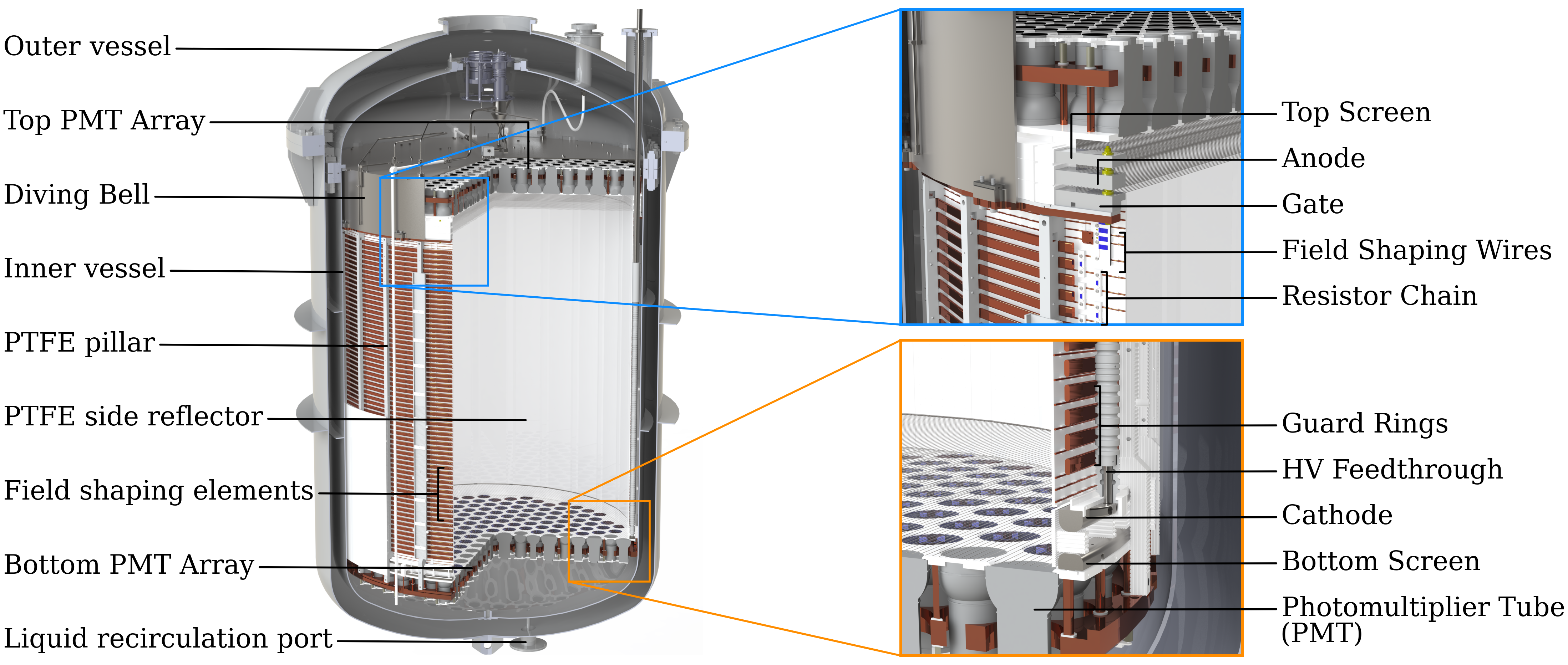}
\hfill
\caption{\label{fig:cad_tpc}CAD rendering of the XENONnT TPC within the cryostat. The zoomed insets show details about the field cage, the electrode stacks at the top and bottom of the TPC, and the implementation of the PMT arrays.}
\end{figure*}

The central detector of XENONnT consists of a LXe target instrumented as a two-phase TPC. It measures both xenon excitation and ionization induced by a particle interacting in the target volume.
This technology is sensitive to low-energy depositions, with ionization signals down to the single electron level~\cite{Aprile:2006kx}. This makes two-phase xenon TPCs ideal for dark matter searches, as demonstrated by the pioneering work of XENON10~\cite{Angle:2007uj} and ZEPLIN-II~\cite{Alner:2007ja} and, more recently, by the results from XENON100~\cite{Aprile:2011dd}, LUX~\cite{Akerib:2012ys}, PandaX-II~\cite{PhysRevD.93.122009}, XENON1T~\cite{Aprile:2017aty}, PandaX-4T~\cite{PandaX:2018wtu}, LZ~\cite{LZ:2022ufs} and XENONnT~\cite{PhysRevLett.129.161805,XENON:2023cxc}.

Energy depositions in two-phase LXe TPCs excite and ionize the xenon atoms, giving rise to signals detected by two arrays of photodetectors installed at the top and bottom ends of the cylindrical TPC.
The prompt, primary scintillation signal (S1) is from photons produced during the de-excitation of the Xe$_2^*$ excimers formed during an interaction. The secondary scintillation signal (S2) is proportional to the number of ionization electrons which did not recombine and were drifted to the gas layer above the LXe target in an electric field across the active target. Once they reach the liquid-gas interface, a stronger electric field (5-10~kV/cm) extracts them into the gas, where they produce proportional electroluminescence.
In a homogeneous electric field, the time delay between the prompt S1 and the delayed S2 is directly related to the length of the electron drift path and thus used to derive the depth ($Z$-coordinate) of the interaction with respect to the liquid-gas interface. The $X$- and $Y$-coordinates are inferred from the distribution of the S2~light on the top-array PMTs. Full 3D~reconstruction facilitates the study of the topology of the interaction (i.e., single- vs multi-site, point-like vs extended tracks) and enables the definition of a fiducial volume with a greatly reduced background level in the analysis~\cite{Aprile:2019bbb}.

The relative sizes of the S1 and S2 signals depend on the energy and nature of the interaction. A particle scattering off a xenon nucleus (e.g., a WIMP or a background neutron) would lead to a short, dense ionization track from the resulting nuclear recoil (NR). The much more abundant backgrounds from intrinsic beta and gamma decays affect atomic electrons, generating electronic recoils (ER's) responsible for ionization tracks sparser than those from NR's and hence relatively less affected by recombination.
The signal ratio~S2/S1 is related to the electron-ion recombination and, hence, to the ionization density generated by the NR or ER. For this reason, this ratio can be used to discriminate between the two, with a typical efficiency as measured in XENON1T of $>$99.7\% to reject ER's while keeping a 50\% NR acceptance~\cite{PhysRevLett.119.181301}. 

\subsubsection{TPC Design}
\label{sec::tpcdesign}

Several design features of the XENON1T detector~\cite{Aprile:2017aty}, a test bed for technical solutions for tonne-scale LXe detectors, were reconsidered and new solutions were implemented for XENONnT. Its TPC was designed with special attention to minimize the amount of plastics (mainly PTFE) needed for structural, insulation, and light reflection purposes. Beyond being a source of neutron background via $(\alpha,n)$-reactions, plastics were measured to be the dominating source of outgassing, degrading the xenon purity and the detector performance. The XENONnT TPC design also minimized the number of individual pieces to reduce the number of joints, surfaces, cavities, bore holes, and threads which can trap air and affect the purity through outgassing and virtual leaks. The uniformity of the drift field was improved compared to XENON1T, mitigating unwanted edge effects and potentially time-dependent charge-up effects~\cite{Aprile:2018dbl}. Finally, in contrast with the hex-etched meshes used in XENON1T, the increased dimensions of XENONnT made it necessary to replace them with parallel wire grids. For the anode and gate electrodes, their mechanical stability was improved by a few transverse wires (see Sec.~\ref{sec::electrodes}). The overall goal of the design was to fit the largest possible TPC into the XENON1T outer cryostat vessel of 1620\,mm inner diameter.

\begin{figure}[t]
\centering 
\includegraphics[width=.49\textwidth]{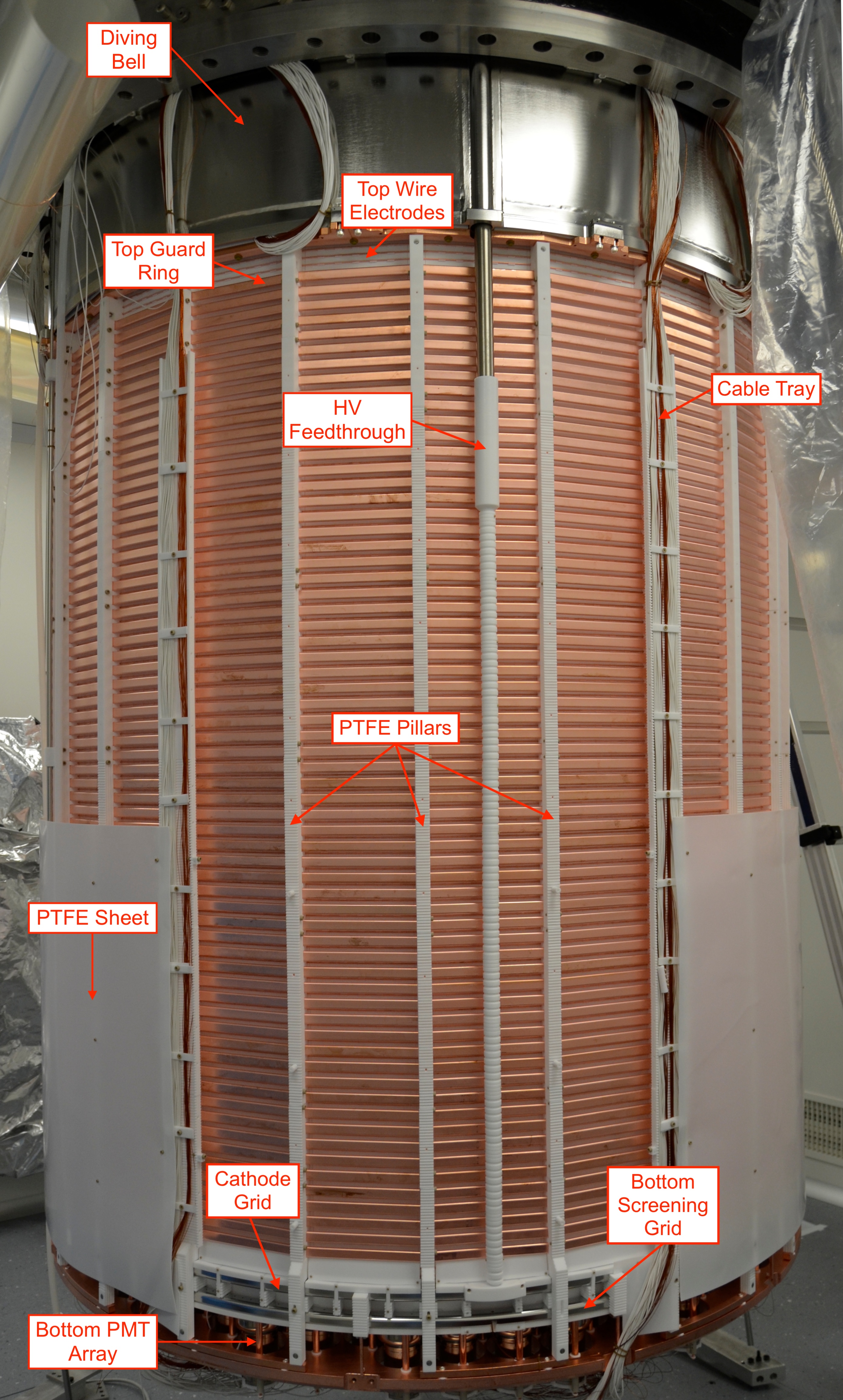}
\hfill
\caption{\label{fig:tpc_photo}The XENONnT TPC suspended from the top flange of the inner cryostat vessel shortly before the assembly was completed within the  cleanroom inside the water tank. One panel of the PTFE sheet surrounding the lower part of the TPC was removed to show how the cathode high voltage feedthrough connects to the cathode electrode.}
\end{figure}

A CAD rendering of the XENONnT TPC is shown in Fig.~\ref{fig:cad_tpc}. The double-wall cryostat contains 8.5\,tonnes of xenon at $-$98$^\circ$C, of which 5.9\,tonnes are inside the TPC active volume. The TPC is suspended from the top flange of the inner vessel and is held together by 24~thin PTFE pillars. The virtually cylindrical active region of the TPC is 1613\,mm high and about 1327\,mm wide (all dimensions in this section refer to the TPC geometry at $-$98~$^\circ$C, i.e., accounting for thermal shrinkage). The cylinder side wall is constructed of 48~interlocking vertical PTFE panels designed to compensate for radial thermal shrinkage while guaranteeing light tightness: 24~panels are fixed to the PTFE pillars and feature a several~mm deep slit on both vertical sides. The additional 24~PTFE panels are placed in between the fixed panels by inserting them into these in an interlocking fashion.  To reduce the mass of PTFE, the thickness of the panels was reduced to 3~mm, the smallest thickness required to ensure opaqueness to xenon scintillation light entering from outside the TPC volume~\cite{Althueser_2020}. The total mass of PTFE used for the TPC is only 128\,kg. The remaining, non-instrumented LXe volume surrounding the TPC with an average thickness of 64\,mm further shields the active volume from radiation from the cryostat. Figure~\ref{fig:tpc_photo} shows the fully assembled TPC suspended from the cryostat vessel.

\begin{figure}[t]
\centering 
\includegraphics[width=.49\textwidth]{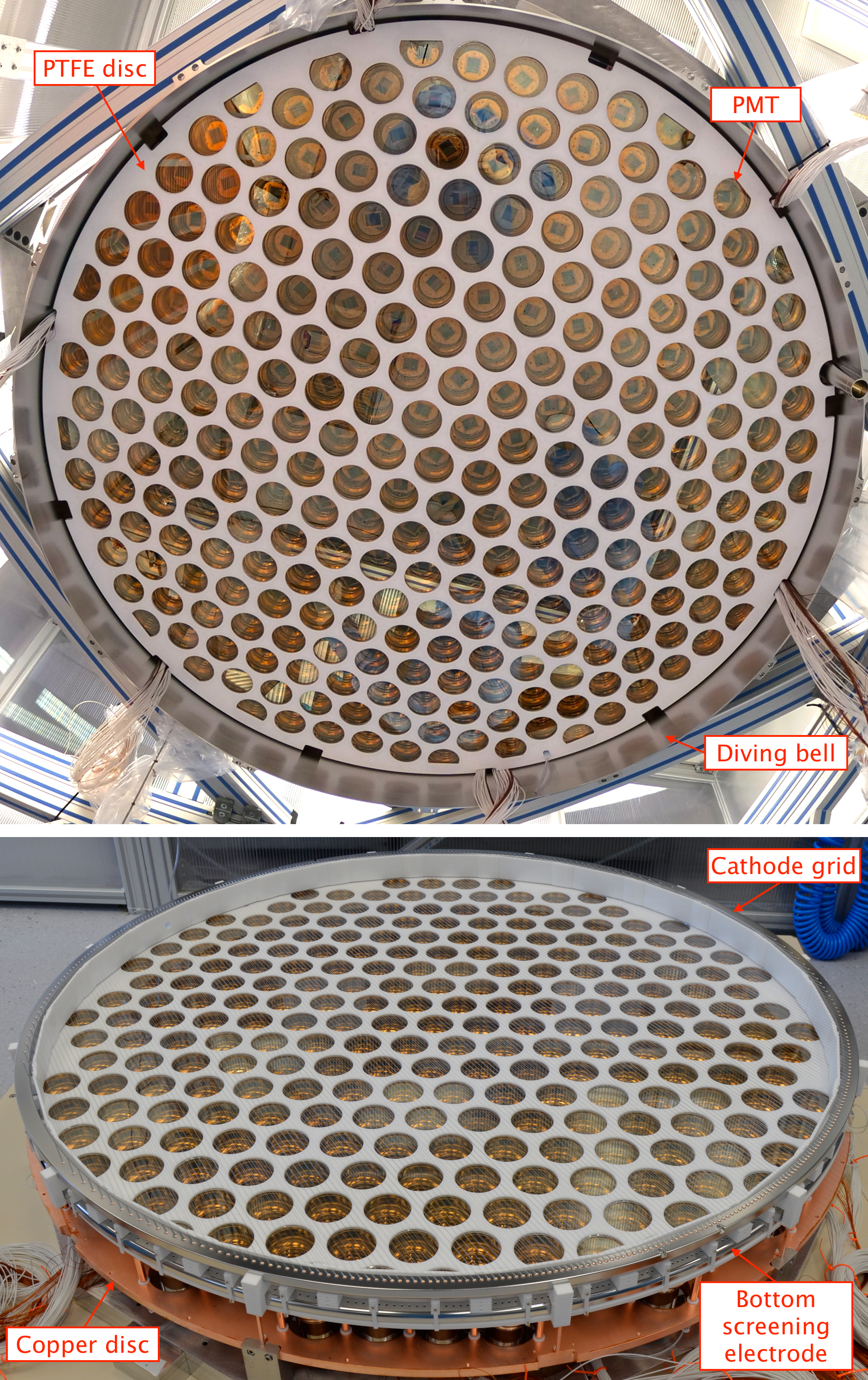}
\hfill
\caption{\label{fig:pmt_array_photo}PMT arrays during TPC assembly: (top) The top array installed inside the diving bell. (bottom) The cathode and bottom screening  electrodes are installed above the bottom PMT array.}
\end{figure}

Two arrays of PMTs detect the S1 and S2 light, with 253 PMTs at the top and 241 PMTs at the bottom (see Fig.~\ref{fig:pmt_array_photo} and Sec.~\ref{sec::pmts} for more details). All PMT bodies in an array are fixed to a single 8-mm-thick PTFE disk filling the gaps between them and acting as a reflector. A single 25\,mm (bottom) or 20\,mm (top array) thick disk made of oxygen-free high conductivity (OFHC) copper serves as a mechanical support for each array, granting stability and flatness during the thermal shrinkage of the PTFE disk. The top PMT array is located inside a diving bell made of 4\,mm-thick stainless steel that isolates the gas in the active volume from the rest of the system. In addition, this allows for pressurization to enable active control of the LXe level. All PTFE surfaces facing the target volume are diamond polished to optimize their reflectivity for 175\,nm xenon scintillation light~\cite{FUJII2015293}.

\begin{figure}[t!]
\centering 
\includegraphics[width=.39\textwidth]{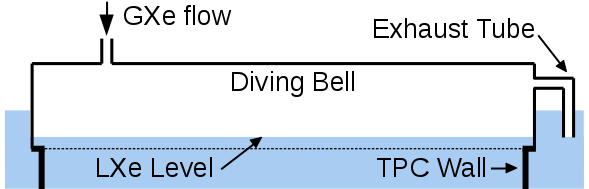}
\hfill
\caption{\label{fig:bell}Illustration (not to scale) of setting LXe level of the TPC with a diving bell: The lower end of the exhaust pipe defines the height of the liquid-gas interface.}
\end{figure}

Five parallel wire grid electrodes and two concentric cages of field-shaping electrodes establish the electric field across the TPC (see Sec.~\ref{sec::electrodes}). The cathode, gate, and anode grid electrodes define the field in the electron drift and electron extraction/multiplication regions, whose lengths are 1486\,mm and 8\,mm, respectively (see Fig.~\ref{fig:cad_tpc}). Two additional screening electrodes are placed 1.5\,mm above the bottom PMT array reflector and 28\,mm below the top PMT array reflector. These are used to reduce the electric fields in front of the PMTs. To avoid an excessively high field below the cathode the bottom screening electrode is located 55\,mm below the cathode, creating a region containing about 200\,kg of LXe from which no S2~signal can be detected as the liberated ionization electrons are drifted to the bottom screening electrode and will not reach the liquid-gas interface. 

Two concentric stacks of field-shaping rings made of OFHC copper are set at incremental potentials via resistor chains from the cathode to the top field-shaping electrode (TFSE) isolated from the gate for better drift field uniformity (see inset in Fig.~\ref{fig:cad_tpc} and Sec.~\ref{sec::electrodes}). A 1\,mm-thick PTFE sheet externally surrounds the lower 640\,mm of the field-shaping electrodes as well as the cathode frame to minimize risks of discharges towards the cryostat which is at ground potential (see Fig.~\ref{fig:cad_tpc} and~\ref{fig:tpc_photo}). The five grid electrodes as well as the topmost field-shaping electrode are independently biased (see Sec.~\ref{sec::electrodes}). The negative HV for the cathode is delivered by the same custom high-voltage feedthrough (see Fig.~\ref{fig:tpc_photo}) used in XENON1T~\cite{Aprile:2017aty}. Kapton-insulated cables~\cite{accuglass26AWG}, connected to the air side by commercial ConFlat flanges equipped with single-ended SHV feedthroughs, deliver the bias to the remaining electrodes. 

\begin{figure*}[t!]
\centering 
\includegraphics[width=.95\textwidth]{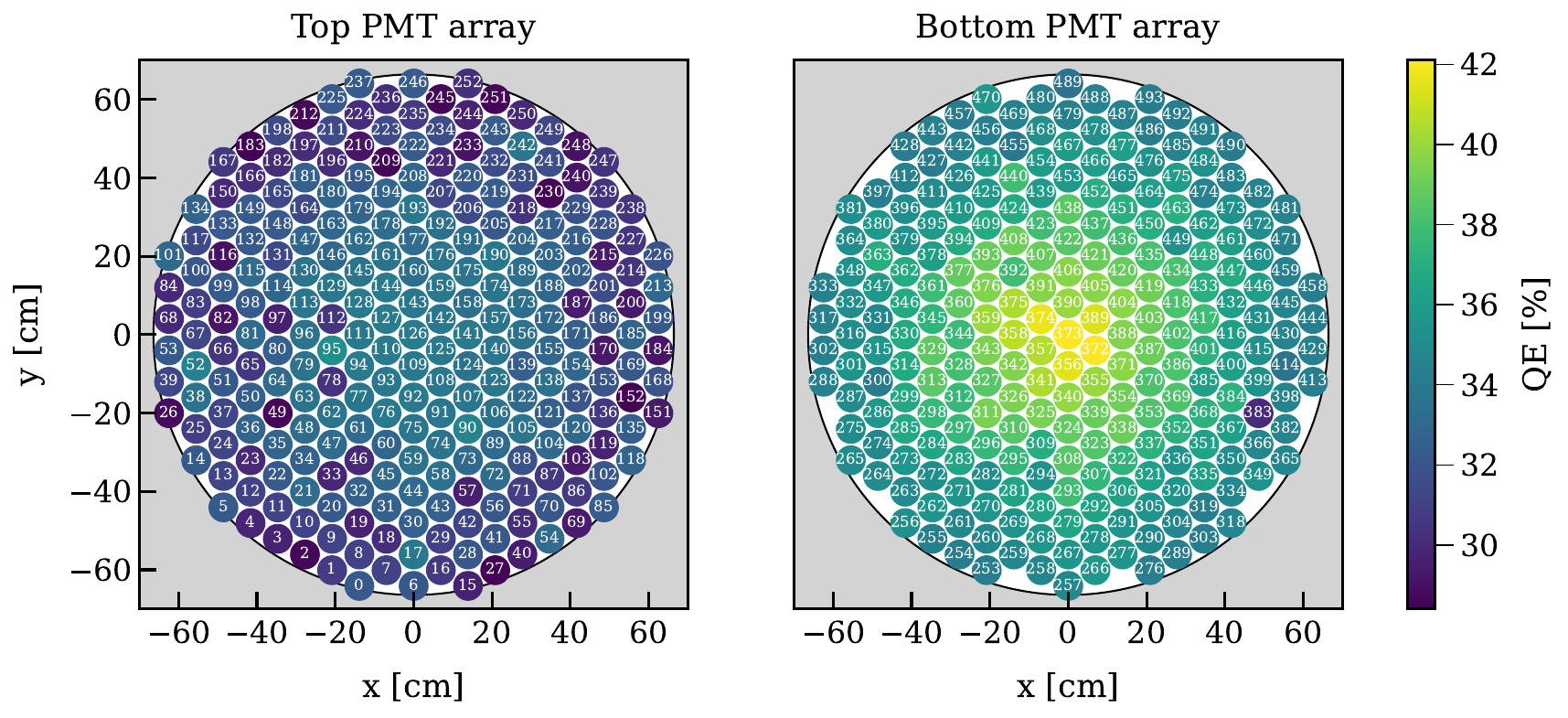}
\hfill
\caption{\label{fig:pmt_array_qe}The Hamamatsu R11410-21 PMTs in the top (253 units) and bottom (241) arrays were distributed according to their room-temperature quantum efficiency (QE, color code). PMT 383 was added just before assembly to replace one of the PMTs which performed worst during testing, i.e., it was not positioned based on its QE.}
\end{figure*}

\begin{table*}[t]
    \centering
    \begin{tabular}{l|l|l|l|l|l|l|l|l|r}\hline
      TPC Electrode   & Frame & Frame & Frame & Frame & Wire & Wire & Optical & Vertical & Design\\
                      &  Thickness &  Shape & ID & OD & Diameter &  Pitch & Transp. & Position & Voltage\\ \hline
     Top Screen    & 15 mm & 24-gonal &  1334 mm & 1408 mm & 216 \mum{} & 5 mm & 95.7\% & $+$36 mm & $-$1.5 kV\\
     Anode            & 18 mm & 24-gonal &  1334 mm & 1408 mm & 216 \mum{} & 5 mm & 95.7\% & $+$8 mm &  $+$6.5 kV\\
     Gate             & 20 mm & 24-gonal &  1334 mm & 1408 mm & 216 \mum{} & 5 mm & 95.7\% & 0 mm & $-$1.0 kV\\
     Cathode          & 20 mm & circular         & 1347 mm & 1395 mm & 304 \mum{} & 7.5 mm & 96.0\% & $-$1486 mm & $-$30.0 kV\\
     Bottom Screen & 15 mm & circular         & 1345 mm & 1395 mm & 216 \mum{} & 7.5 mm & 97.0\% & $-$1541 mm &  $-$1.5 kV\\ \hline
    \end{tabular}
    \caption{Geometrical properties and design voltages of the five grid electrodes. The frame thickness refers to the height of its vertical cross-section. The inner and outer diameters (ID and OD) for the 24-sided polygon electrodes should be viewed as the diameters of the inscribed and circumscribed circles. The nominal vertical positions of the wire planes (intended as the plane crossing the center of the wires) are given with respect to the gate electrode once the TPC is cooled down to LXe temperature.}
    \label{tab:electrodes}
\end{table*}

The position of the liquid-gas interface is mechanically set by a vertically-adjustable gas-exhaust tube, a 1/4" stainless steel tube running along the diving bell's sidewall. Its upper end is connected to the gas in the diving bell while the lower one ends in the LXe volume outside the TPC, see Fig.~\ref{fig:bell}. Pressurizing the bell by a GXe flow sets the level of the liquid-gas interface inside the bell to the lower end of the exhaust tube. Its vertical position can be controlled remotely by a linear motion feedthrough and was set to keep the liquid-gas interface stable at 5.1\,mm above the gate electrode. This value refers to the nominal gate position,
i.e., the surface defined by the thick electrode support frames. In reality, under the combined effect of gravity and the electric field, the electrode wires deform, leading to a variation of this distance as a function of the $X$-$Y$-coordinate that is corrected for in the offline analysis (see Sec.~\ref{sec::S2respperf}). To monitor the liquid level, four parallel-plate capacitive level meters were installed along the circumference of the TPC. These can monitor the level with a 0.2\,mm precision over a dynamical range of 10\,mm. The same level meters are used to measure the potential tilt of the TPC and hence the orientation of the TPC electrodes' nominal planes with respect to the liquid-gas interface. The tilt can be adjusted at 50\,\mum{} precision from outside the water shield by a leveling system connected to the cryostat.  Analysis of the gas gap using the width of single electron signals confirms good leveling, with an estimated tilt of only $(1.8 \pm 0.1)\times 10^{-4}$\,rad. 

During the filling of the cryostat with liquid xenon, the liquid level is monitored with two capacitive tube-in-tube level meters that run alongside the TPC.

All cables (for electrodes, PMTs, and sensors) are routed via two vacuum-insulated double-walled pipes, which connect the cryostat to the air side through the water shield (see Fig.~\ref{fig:globalscheme}). The cable pipe~1 is part of the large multi-purpose cryostat pipe that connects the cryostat top dome to the third floor of the XENON service building and also contains other cryogenic lines. This large pipe, including all the cables inside,  was re-used from XENON1T. The second cable pipe~2 was newly installed on a separate port on the cryostat top dome to accommodate the remaining cables. A constant xenon gas extraction from these pipes towards the cryogenic system avoids potential back diffusion of radon emanated by the cables and components at a higher temperature (see gas xenon flows in Fig.~\ref{fig:globalscheme}).

\subsubsection{TPC Photomultiplier Tubes}
\label{sec::pmts}

The top and bottom arrays include 253 and 241 Hamamatsu R11410-21 low-background cryogenic PMTs~\cite{R11410-21}, respectively. These were developed jointly by Hamamatsu and the XENON collaboration~\cite{Baudis:2013xva,Barrow:2016doe,Antochi_2021}. The PMTs were selected from a pool of 520 units, 153 recovered from XENON1T (out of~248), and 367 new units. All the available PMTs were stress-tested in liquid xenon conditions to catch additional potential failure modes, as XENON1T had seen almost 10\% of PMTs failing during its operation, despite having been stress-tested in cold nitrogen gas. Out of the 367 new PMTs tested, 15 showed a high after-pulsing rate, and 11 emitted light at a high rate, and were not used for XENONnT~\cite{Antochi_2021}). The best performing 494~units were identified and installed. The installed PMTs have an average quantum efficiency (QE) of 34.1\% measured by Hamamatsu at 20$^\circ$C for 175\,nm light and an average collection efficiency of 90\%~\cite{Antochi_2021}. 

To maximize coverage, the PMTs on each array are distributed in a hexagonal, maximum-packing pattern as shown in Fig.~\ref{fig:pmt_array_qe}. Optical simulations of the position reconstruction using the S2~signal distribution on the top array showed that the close-packing layout is more resilient to reconstruction bias induced by potential failing PMT channels compared to a radial pattern. The PMTs with the highest QE were installed in the bottom array to maximize the light collection efficiency (see Fig.~\ref{fig:pmt_array_qe}). In both arrays the higher the QE the closer the PMT was placed towards the center, while taking care that PMTs showing sub-optimal performance of any type were not clustered in space. In the top array, further care was put to avoid placing sub-optimal PMTs near the edges, where a failure would compromise our fiducialization capabilities.

The PMTs are individually biased~\cite{caenPMT} and equipped with high-voltage dividers installed on the Cirlex\textsuperscript{\textregistered} printed voltage-divider circuit boards developed for XENON1T~\cite{Barrow:2016doe}. These bases feature linearity up to a signal current of about 30~mA at 200~Hz signal rate. The PMT high-voltage cables~\cite{kaptonUHV} and signal cables~\cite{RG196-1,RG196-2} reach the service building through the two cable pipes introduced in Sec.~\ref{sec::tpcdesign}. The cables are potted into ConFlat flanges~\cite{rhseals} with leak rates below 10$^{-8}$\,mbar\,l/s. The high-voltage return lines and the ground lines of the PMTs (coaxial shields) are kept separated until they reach the DAQ electronics. PMT single-photoelectron calibrations are performed by injecting blue light (465~nm) into the active TPC volume through a network of optical feedthroughs and fibers.

\begin{figure}[t]
\centering
\begin{subfigure}{.999\columnwidth}
\centering
\includegraphics[width=\columnwidth]{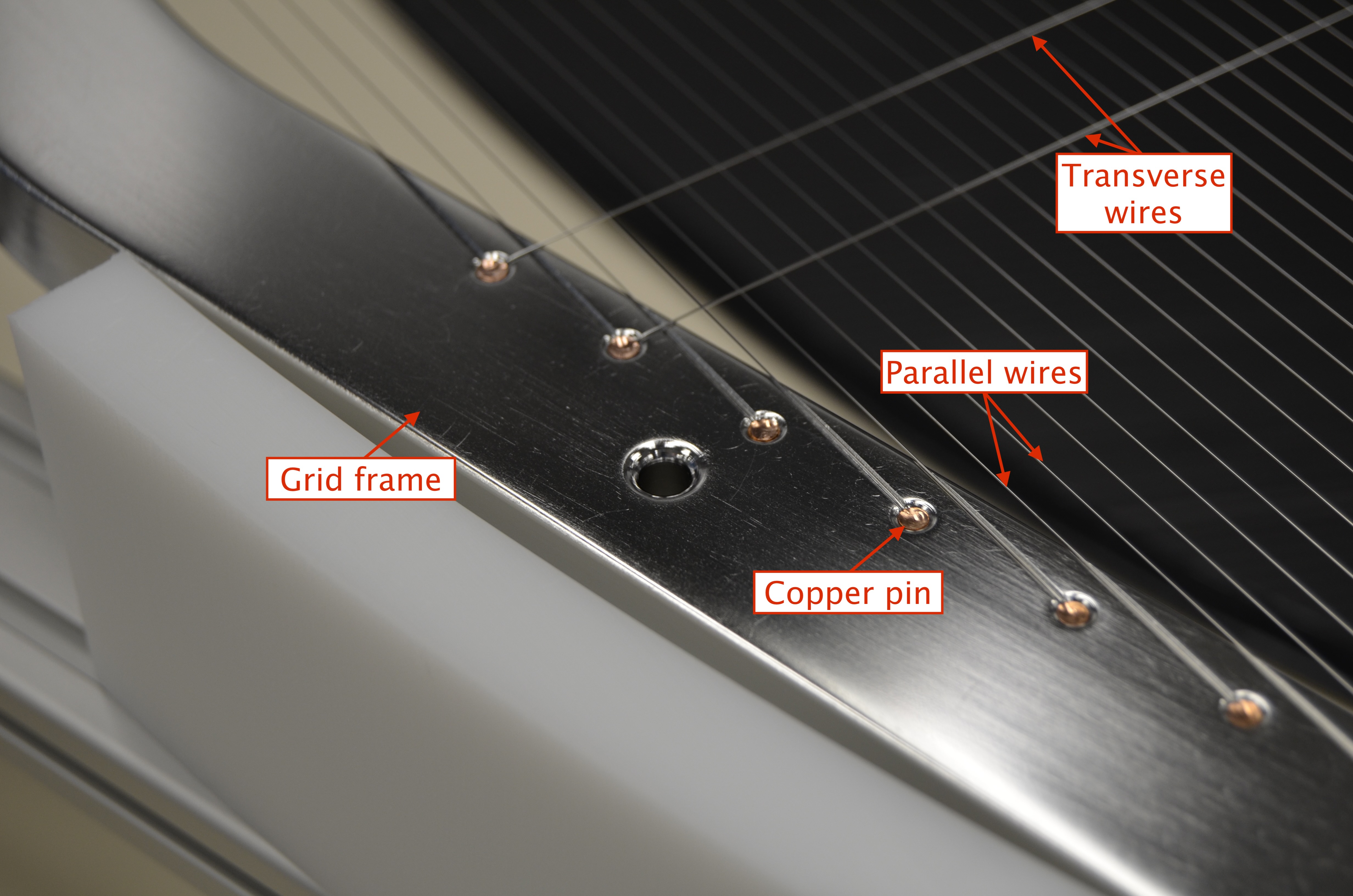}
\end{subfigure}\hfill
\begin{subfigure}{.999\columnwidth}
\centering
\includegraphics[width=\columnwidth]{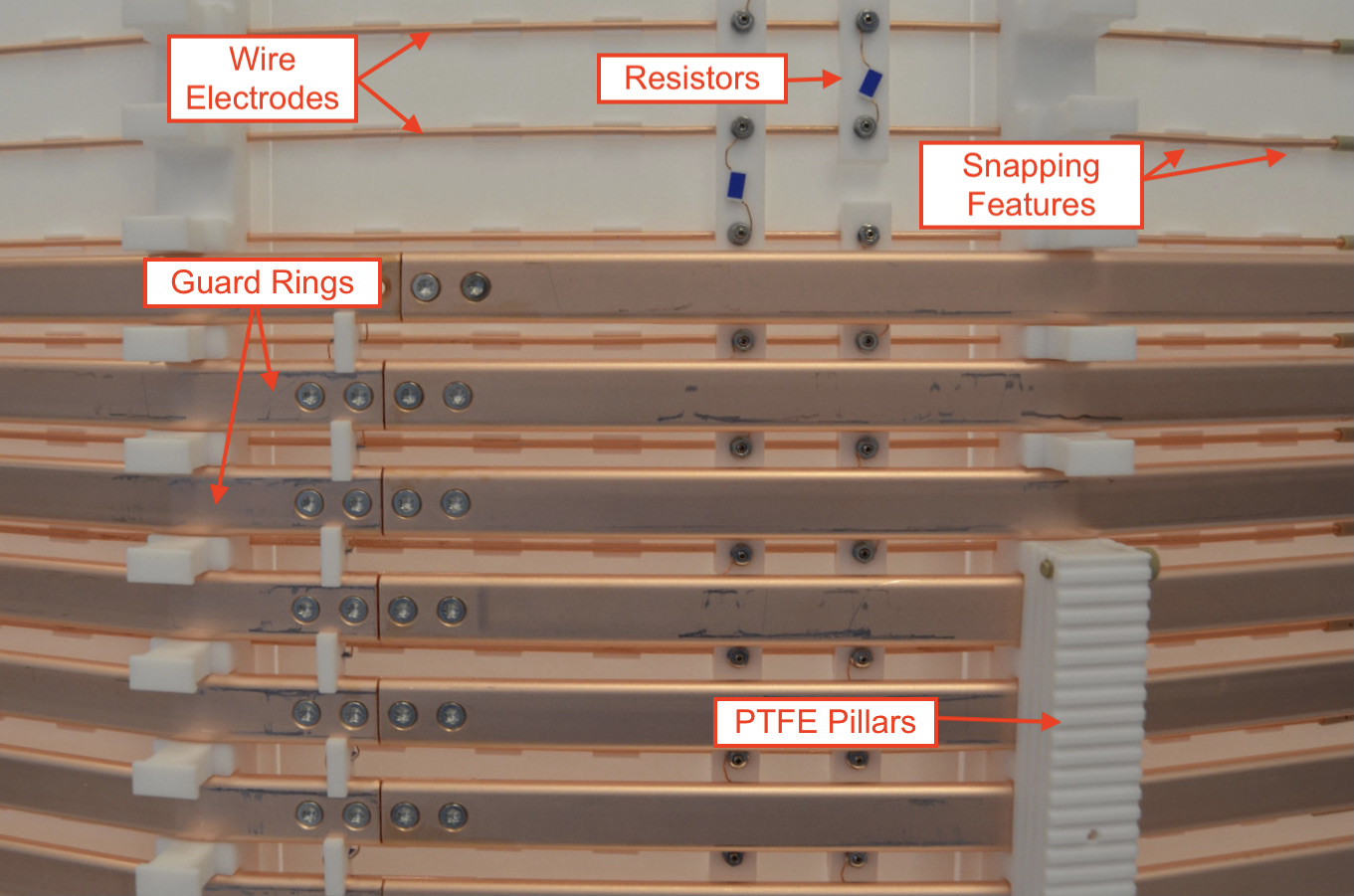}
\end{subfigure}
\caption{\textbf{(top)} The electrode wires are individually fixed by copper pins to the electrode frames (here: anode). Two transverse wires, installed to reduce sagging, are shown as well. \textbf{(bottom)} Detailed outside view of the TPC field-shaping elements during installation: the massive guard rings (15\,mm height) and the thinner wire field electrodes (2\,mm diameter) are installed at half pitch at different distances from the 3\,mm thick PTFE reflectors. The guard electrodes consist of two pieces fixed together by two M3~SS bolts at each junction. The other two bolts are used to fix the guard-ring resistor chain (installed within the guard rings). Also visible are the 5\,G\textohm~resistors on PTFE supports connecting the wire field electrodes.}
\label{fig:fieldcage}
\end{figure}

\subsubsection{TPC Electrodes}
\label{sec::electrodes}

\paragraph{Electrodes:}
The electrodes are made of parallel wires stretched across massive frames which were machined from single slabs of SS304 stainless steel (see Tab.~\ref{tab:electrodes} for geometrical details). The material was selected for its low radioactivity and electropolished after the production of the frames (see Sec.~\ref{sec::materials}). The geometry of the frame for each grid was optimized with 3D mechanical simulations to mitigate deformations under the effect of the wire tension while minimizing the amount of material. All edges were rounded with radii  $\sim$1\,mm to avoid local high electric fields and potential discharges.

Annealed SS316 stainless steel wires~\cite{cfw} were selected for their smooth surface to improve high-voltage stability. Wire diameter and pitch for each electrode were chosen to maximize the optical transparency (as light generated in the active LXe target or the S2 region has to traverse at least two electrodes) while ensuring a good uniformity of the electric field. Every wire is fixed individually to the frame by two OFHC copper pins (1.98\,mm diameter) that are pushed, together with the tensioned wire, into holes of 2.1\,mm diameter machined along the frames (see Fig.~\ref{fig:fieldcage}, top). The pins are fully contained in the frames but can be removed to allow for wire replacement. The grid electrode assembly procedure applies a pre-deformation of the frame achieved by a push/pull-system of 12~stations arranged radially around the frame. This pre-deformation guarantees uniform tension in all wires that are installed on an electrode. The wires are fixed on one side of the frame, tensioned with a precisely controlled force between 3 and 4\,N (according to their position), and then fixed on the other side. Grooves of 0.36\,mm depth in the frame guide the wires. All frames and wires were cleaned prior to assembly (see Sec.~\ref{sec::materials} and Ref.~\cite{XENON:2021mrg}). To mitigate potential electron and photon emission, they were passivated in a 7\% citric acid solution similar to~\cite{Tomas:2018pny,Linehan:2021qnb}.

Mechanical and electric testing of all electrodes was performed in air and in LXe inside the XENONnT cryostat during the commissioning of the cryogenic and purification systems. Particular emphasis went into ensuring that the wires were able to withstand vibrations that may occur during transportation and installation as well as the cool-down to LXe temperature. Two and four additional wires of 304\,\mum{} diameter were installed transversely to the others on the gate and anode electrode, respectively (see Fig.~\ref{fig:fieldcage}, top) to counteract deformation of the electrode planes under electrostatic forces. In this configuration, the nominal gate-anode distance of 8\,mm reduces to a minimum of $\sim$7\,mm, while at LXe temperature and at the design potentials (see Tab~\ref{tab:electrodes}). The impact of the presence of transverse wires on S2 signal formation is described in Sec.~\ref{sec::S2respperf}.

\paragraph{TPC Field Cage:}
Two sets of field-shaping (FS) electrodes, located outside the PTFE reflector walls, surround the active volume (see Fig.~\ref{fig:fieldcage}, bottom). Together with the gate and cathode electrodes, they define the electric field in the drift region. Studies based on~2- and 3-dimensional COMSOL Multiphysics electrostatic simulations showed that the double cage structure improves the uniformity of the drift field and maximizes the active volume and it helps to avoid charge-up of the PTFE walls of the TPC~\cite{XENONnT:2023dvq}. 

The outermost set consists of 64~OFHC copper electrodes (hereafter referred as ``guard rings''), with radii 10.7\,mm larger than the PTFE reflector wall's inner radius. Each guard ring is shaped as a 24-sided polygon, formed by two sections screwed together with M3 stainless steel bolts (see Fig.~\ref{fig:tpc_photo} and~\ref{fig:fieldcage}, bottom) and features a 15\,mm$\times$5\,mm cross-section with rounded corners (2.5~mm radius).  The rings are equally spaced and kept in place by notches in the PTFE pillars. The vertical center-to-center distance (pitch) between two adjacent guard rings is 21.6\,mm at LXe temperature. Due to space constraints, the bottom edge of the lowest ring is located 33\,mm above the cathode wire plane while the topmost one is positioned 73\,mm below the gate plane as shown in the two insets of Fig.~\ref{fig:cad_tpc}. 

A second set of 71~concentric thinner copper electrodes (hereafter referred to as wire field electrodes) is installed between the guard ring cage and the PTFE reflector walls. Each wire electrode is made of a 2\,mm OFHC copper wire shaped as a 24-sided polygon. The wire field electrodes, touching the outer surface of the PTFE reflector panels, are held in place by five elevated notches machined into each PTFE panel. Holes of 0.25\,mm diameter were drilled through the panel right in the middle of the notches (underneath the snapped-in copper wire) to allow for the removal of charge potentially accumulating on the internal surface of the PTFE reflectors. The wire field electrodes have a 21.6\,mm pitch and are vertically staggered, with respect to the guard rings, by half pitch (see Fig.~\ref{fig:fieldcage}, bottom). However, in the regions without guard electrodes, right above the cathode and below the gate electrode, their distance is reduced to 11\,mm to mitigate field leakage. Since both the guard and wire field electrodes are anchored to PTFE components (see Fig.~\ref{fig:fieldcage}, bottom), both field cages shrink coherently when cooled down, maintaining the desired geometrical scaling. 

Two voltage divider networks, composed of 5\,G\textohm\ resistors~\cite{ohmite}, run laterally to the active volume and establish a vertically decreasing potential for the wire electrode and the guard ring cages. In each cage the voltage steps between successive electrodes, and hence the different resistor combinations between successive rings, were optimized via electrostatic simulations, ensuring maximal field homogeneity. In addition, regions in the TPC where charge is not transported towards the anode are minimized for a very wide range of cathode bias potentials between $-$10\,kV and $-$30\,kV. The first resistor connects the cathode to the wire electrode chain. The topmost wire electrode is not connected to the gate grid and its voltage can be set independently. The second voltage divider, serving the guard rings, couples at the top and bottom to the wire electrode divider and connects the topmost and bottommost rings to the corresponding wire field electrodes (located 58.4\,mm below the gate and 29.5\,mm above the cathode). 
The possibility to independently bias the topmost wire electrode allows for optimizing the field during a run in case the actual voltages differ from the design values. 
COMSOL electrostatic simulations indicate that, at the design voltages of Tab.~\ref{tab:electrodes} and setting a potential of $-$0.95\,kV on the top field-shaping wire, the drift field between the cathode and the gate is expected to have an average strength of E$_\mathrm{d}$=191\,V/cm with an RMS spread of only 1\%~\cite{XENONnT:2023dvq}.

\subsubsection{Detector Construction Materials and Cleanliness}
\label{sec::materials}

\paragraph{Material Selection:} 
A screening campaign was performed to determine the bulk radioactivity and the radon emanation rate of all materials and components to be  used in the TPC or in close proximity to it~\cite{XENON:2021mrg,XENON:2020fbs}.  

Gamma-ray spectrometry and inductively coupled plasma mass spectrometry (ICP-MS) were performed at several screening facilities~\cite{Heusser:2006,Baudis:2011am,Heusser:2015ifa,vonSivers:2016xpo,Garcia:2022jdt}. Raw materials (stainless steel, OFHC copper, PTFE, and Torlon\textsuperscript{\textregistered}), the Hamamatsu PMTs~\cite{XENON:2015ara}, and various smaller components were screened and the most radiopure options were selected. The radioactivity level of the copper and plastic materials is similar to that of XENON1T~\cite{XENON:2017fdb}, while the stainless steel for the cryostat, TPC electrodes, and diving bell is on average slightly more radioactive. A Monte Carlo study was performed in parallel to the screening campaign to inform decisions about the TPC design and the material selection~\cite{XENON:2020kmp}. It showed that the impact of the stainless steel components is marginal compared to the dominant background sources at low energies such as radon-induced ERs.

To achieve the target $^{222}$Rn concentration of 1\,\unitRad{}~\cite{XENON:2020kmp}, only materials with a low radon-emanation rate were used for components in contact with the xenon. The materials were selected through radon emanation measurements performed 
by electrostatic radon monitors or miniaturized proportional counters for alpha detection~\cite{Zuzel:2009}. 

\paragraph{Cleanliness Procedures:}
To further reduce their radon emanation, all components but PMTs went through a two-step cleaning process that was executed inside a class ISO~6 cleanroom above ground at LNGS. To remove residues from the manufacturing process, the components were first treated in an ultrasonic bath filled with a solution of water and detergents selected according to the material~\cite{Bruenner:2020arp}. 
After rinsing with demineralized water, the components were subjected to a second stronger cleaning in an acidic solution (different solutions for different materials) to remove traces of the $^{222}$Rn progenitor $^{226}$Ra from the surface. For components facing the active xenon volume the treatment also reduced long-lived radon daughters, like $^{210}$Pb and $^{210}$Po,  plated out on the surfaces which could contribute to the background ~\cite{Aprile:2019dme}. After cleaning, all components were stored in boxes filled or flushed with dry nitrogen gas to avoid re-contamination. 

The TPC was fully assembled in the same cleanroom on its assembly support structure. Whenever possible, its active volume was flushed with dry clean nitrogen gas to minimize its exposure to air. The final assembly of the components, some of which like the PMT arrays and grids were pre-assembled, took about 14~days. Subsequently, the TPC was packed in a radon impermeable Mylar bag. It was transported to the underground laboratory with a low-vibration truck and stored in a pre-cleanroom area inside the water tank, where the outer surfaces of the bag were intensively cleaned. Finally, the TPC was separated from its assembly support structure and lifted into an ISO~6 hardwall cleanroom constructed within the water tank around the top domes of the cryostat vessels (see Sec.~\ref{sec::cryogenics}). After the TPC was connected to the dome of the cryostat, the PMT signal and HV cables were wired. Throughout the operation, which took 7~days, the active TPC volume was continuously flushed with nitrogen gas and the TPC was bagged whenever possible. The inner cryostat vessel was then closed and evacuated.

\begin{figure*}[t]
\centering 
\includegraphics[width=.98\textwidth]{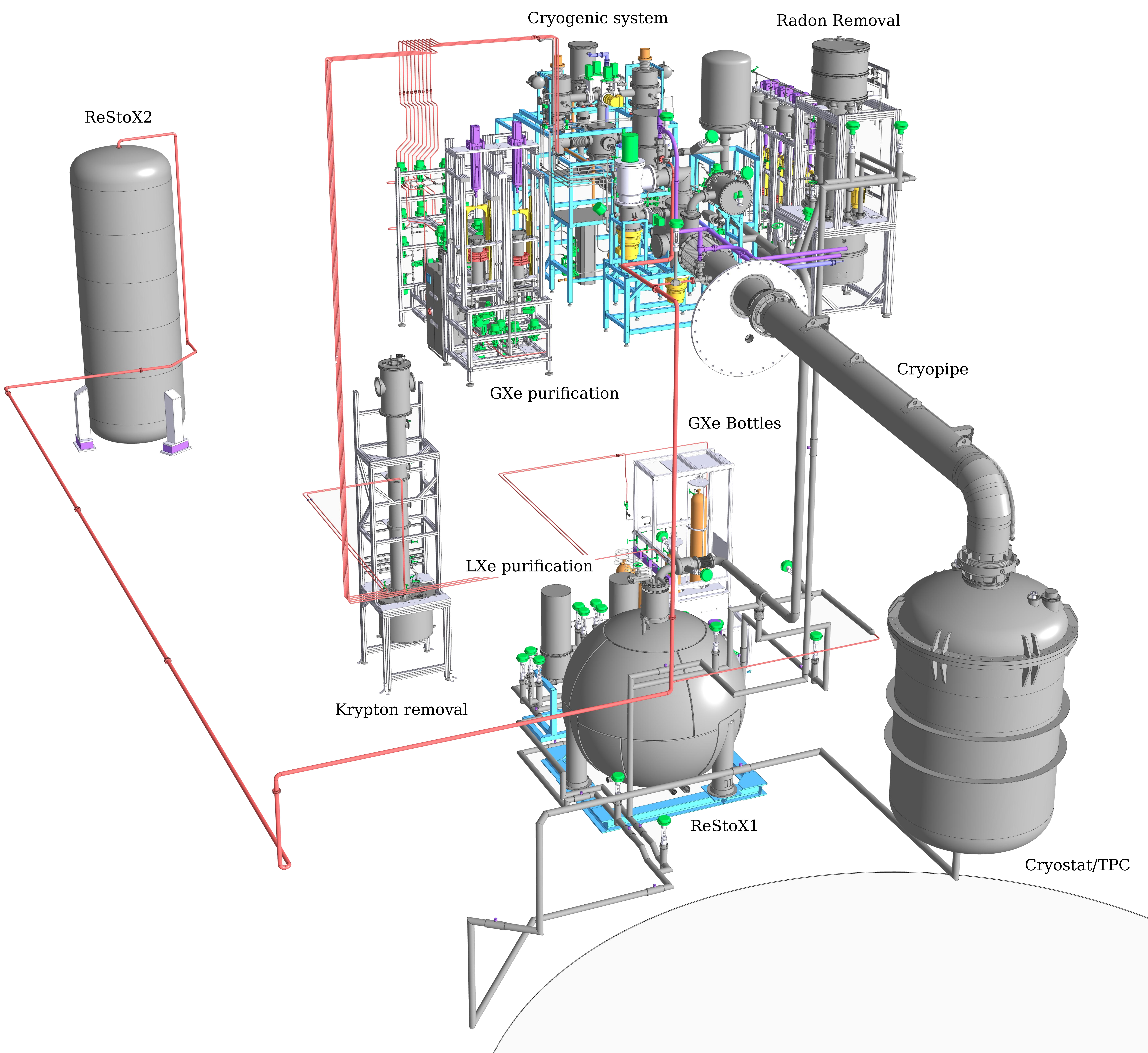}
\hfill
\caption{\label{fig:cad_cryo} CAD rendering of the xenon handling system in XENONnT. It includes the cryostat hosting the LXe TPC, the cryogenic system used to cool the xenon (CRY), the liquid (LXe-PUR) and gaseous purification (GXe-PUR) systems for electronegative impurity removal, the cryogenic distillation columns for krypton (Kr-DST) and radon removal (Rn-DST), ReStoX1 and ReStoX2 for LXe storage, filling and recovery, and the gas bottle rack for adding xenon gas into the system.}
\end{figure*}

\subsection{Xenon Handling}
\label{sec::xehandling}

A CAD rendering of the xenon handling system is shown in Fig.~\ref{fig:cad_cryo}. Substantial upgrades with respect to XENON1T are the LXe purification system, a novel radon removal system, and the second large xenon storage unit (ReStoX2).

\subsubsection{Cryostat and Cooling System}
\label{sec::cryogenics}

\paragraph{Cryostat:}
The TPC is housed inside a vacuum-insulated cryostat composed of two SS304 vessels holding the xenon (see Fig.~\ref{fig:cad_tpc}). The outer vessel (OV) was re-used from XENON1T and refurbished. It keeps the same torispherical head (with the addition of an extra port) but presents an extended cylindrical shell, obtained by welding an extra section. The new OV has an internal diameter of 1620\,mm and a height of 3001\,mm. The inner vessel (IV), 1460\,mm in diameter and 2666\,mm high, was newly fabricated. Both vessels have a wall thickness of 5\,mm. A spring-energized metal seal \cite{helicoflex} is used to close the IV head and shell flanges to ensure vacuum integrity during cold operations. When operated as a cryostat, the OV volume is actively evacuated while 30~layers of superinsulation, surrounding the IV, further reduce the influx of heat.
The IV and OV heads are connected by three massive brackets. The cryostat is suspended in the center of the water tank on three M24 stainless steel rods connecting it to the support structure.

As for XENON1T, the torispherical heads of the vessels are equipped with a central port coupled to a large vacuum-insulated cryogenic pipe that contains the return line from LXe PUR, the gaseous purification system inlet/outlet lines, and the pipe carrying the PMT signal and high-voltage cables (see Fig.~\ref{fig:cad_cryo}). The cryostat head has two additional ports for the cathode high-voltage feedthrough and the motion feedthrough to control the liquid level. A new port was added at the bottom of the cryostat, allowing for the installation of a vacuum-insulated line connecting the cryostat to the inlet of the liquid purification system (see Sec.~\ref{sec::LXe_pur}).

\paragraph{Cooling System:}
The cryogenic control system of XENONnT is largely identical to XENON1T. It is based on the ``remote cooling'' concept introduced for XENON100~\cite{Aprile:2011dd}, with the 6\,m long vacuum-insulated cryogenic pipe connecting the cryostat to an external cooling station located in the XENON building~\cite{Aprile:2017aty}. Heat influx is compensated by a system of two redundant pulse tube refrigerators~\cite{Iwatani} (PTRs), each providing 250\,W of cooling power. Stable temperature regulation is achieved by controllers~\cite{lakeshore} that measure the temperature at a copper block directly coupled to the PTR and supply power to the attached heaters. In case of a short term (few days) power failure, or PTR  equipment failure, the pressure is maintained below the cryostat's maximum working pressure by a battery-powered backup LN$_2$ cooling system composed of a temperature controller and a GN$_2$ proportional valve. In the event of a long term power failure or a complete power loss, the backup LN$_2$ cooling system automatically switches to a mode where it can operate without any electricity and provides a fixed cooling power. 

\subsubsection{Xenon Storage and Recovery}
\label{sec::storage}

The XENONnT xenon handling system has two storage units: ReStoX1 and ReStoX2. They can store the full inventory of xenon, and can rapidly recuperate xenon from the TPC in case of emergency.

ReStoX1 was designed and fabricated for XENON1T~\cite{Aprile:2017aty}. It consists of a vacuum-insulated stainless steel sphere installed on the ground floor of the XENON service building and it is capable of storing up to 7.6\,tonnes of xenon from cryogenic temperatures up to room temperature, as a supercritical fluid (see Fig.~\ref{fig:cad_cryo}). 

ReStoX2 can be operated alongside ReStoX1. It is rated up to 60\,bar and can hold 10\,tonnes of xenon at room temperature, more than the total XENONnT inventory. It was designed not only to expand the storage capacity to the required amount, but also to complement the operation capabilities of ReStoX1.

Xenon recovery in XENONnT makes use of both ReStoX units. Recuperation in liquid phase is possible by directly guiding the LXe into ReStoX1 through the vacuum-insulated pipe connected at the bottom port of the cryostat. The xenon exceeding the ReStoX1 capacity can be transferred in gas phase to ReStoX2 through a direct connection. Recuperation of gaseous xenon can be achieved through cryogenic pumping, by freezing the xenon into ReStoX2. For rapid recuperation of the gaseous xenon from the cryostat, a valve is opened and liquid nitrogen starts flowing through a large parallel plate heat exchanger, installed inside the ReStoX2 tank. Due to its 100\,m$^2$ exchange surface, the xenon can be recuperated with transfer rates exceeding $1$\,t/hour. ReStoX1 can be used to depressurize the cryostat in the case of a moderate excess in heat input such as a loss of the cryostat's vacuum insulation. ReStoX2 can also be used for that purpose, handling even larger heat input scenarios. 

The XENONnT cryostat was filled by recondensing the xenon gas (instead of injecting the liquid phase directly) in order to run the full gas inventory through the GXe purification system to suppress electronegative impurities as well as potential traces of tritium~(T) in the form of HTO or HT contained in the stored gas (see Sec.~\ref{sec::GXe_pur}). 
For this procedure, the LXe in ReStoX1 was evaporated, passed through the hot getter, and recondensed again (see Sec.~\ref{sec::cryogenics}).

\subsubsection{Removal of Electronegative Impurities}
\label{sec::xenon}

The removal of electronegative impurities (mainly O$_2$) is essential for operating the $\sim$1.5\,m tall TPC. Given the increase of the target mass, the GXe purification system of XENON1T was complemented with a novel LXe purification system in order to reach a high purity on relatively short timescales~\cite{XENON:Plante_2022}.

\paragraph{GXe Purification:}
\label{sec::GXe_pur}
The system purifying the xenon gas is mainly unchanged with respect to XENON1T~\cite{Aprile:2017aty}. During standard operations the LXe is extracted from the cryostat, outside of the TPC, through a line running inside the large cryogenic pipe, evaporated through a combination of pipe-in-pipe and plate heat exchangers and fed into the gaseous purification system, located on the third floor of the XENON service  building (see Fig.~\ref{fig:cad_cryo}). Impurities are removed from the xenon in two parallel branches, each equipped with a high-temperature rare-gas purifier 
(getter)~\cite{PS4-MT50-R}. In view of potential traces of tritium observed in XENON1T~\cite{PhysRevD.102.072004}, the Hydrogen Removal Units (HRUs) of the getters were fully regenerated for XENONnT. Except for one unit used in the radon distillation column (see Sec.~\ref{sec::rn}), all GXe pumps (QDrive) employed in XENON1T were replaced in XENONnT by custom-developed magnetically-coupled piston pumps~\cite{Brown:2018uya}. One single pump (having a second as a backup) drives a maximum flow of 50\,SLPM per purification branch, the highest flow at which the gas purifier efficiency remains within its specification. In addition, these pumps have a significantly lower radon emanation rate as demonstrated by a test performed at the end of XENON1T operations~\cite{XENON:2020fbs}. The system is typically operated at $\sim$80\,SLPM, equally split among the two purifier branches. The gas, once purified, is liquefied using the pipe-in-pipe and plate heat exchanges and returned to the cryostat. 
The performance of the system is described in more detail in Sec.~\ref{sec::commissioning}. The gas purification system is connected to all other relevant systems, which for example facilitates the injection of radioactive sources for calibration and the removal of hydrogen.

\paragraph{LXe Purification:}
\label{sec::LXe_pur}
In the LXe purification system, the liquid xenon is extracted from the vacuum-insulated port at the bottom of the cryostat (see Fig.~\ref{fig:cad_cryo}). It is further cooled down by a  LXe-LN$_2$ heat exchanger that counterbalances the heat load for a null net heat input on the cryostat and assures appropriate sub-cooling to allow for pumping of the liquid xenon. It is then pushed through one of two parallel purification branches. The purified LXe is returned into the cryostat through a LXe line contained in the cryogenic pipe. The system is equipped with a clean cryogenic liquid pump with a low heat influx~\cite{barber} to establish a volumetric flow of 1-4\,LPM (about 4-16\,tonnes/day), with a redundant pump installed in parallel. The liquid xenon is purified by custom-designed vacuum-insulated purifiers containing SAES St707 getter pills to remove O$_2$ and other electronegative traces during normal operation of the system. St707 was selected for its low radon emanation rate. During initial operations of the system, however, when the outgassing rate and O$_2$ concentrations are relatively high, an Engelhard Q-5~purifier~\cite{q5} is used instead, despite its higher radon emanation, to rapidly improve the purity to a level where the much more expensive St707 purifier can be used effectively. To enable continuous operations, pump and purifier units are enclosed in separate vacuum vessels allowing for independent pump maintenance and the regeneration/substitution of the purifiers. The liquid purification system can purify the entire XENONnT LXe target to $<$0.1 ppb\,O$_2$-equivalent concentration in about 7\,days. The measured performance is reported in Sec.~\ref{sec::commissioning}. 

\paragraph{Purity Measurements:}
The purity of liquid xenon in the context of TPCs, is routinely expressed in terms of electron lifetime $\tau_e$. This is the time that a cloud of electrons would have to travel through the medium before being reduced by a factor~$e$ due to attachment to electronegative impurities. $\tau_e$ can be measured by studying the amplitude of the S2 signals as a function of drift time for monoenergetic interactions within the active volume. To allow for independent measurements a dedicated purity monitor was installed in-line with the purification system, sampling the LXe before or after the purifiers. The monitor design follows the concept described in Ref.~\cite{AMERIO2004329}. A UV light pulse from a xenon flash lamp~\cite{L7685_UV} shining on the monitor cathode releases electrons, which drift across 20 cm through LXe to the anode. The electron lifetime~$\tau_e$ is then determined as
\begin{equation}
\tau_e = -\frac{\Delta t}{\ln{(Q_a/Q_c)}}, 
\end{equation}
with $\Delta t$ being the drift time of the electrons from the cathode to the anode, and $Q_a$ and $Q_c$ the charge measured on the anode and cathode, respectively.
This nearly instantaneous measurement is accurate for the electron lifetime range from 25\,\mus{} to about 30~ms. At longer lifetimes, the short length of the monitor limits its accuracy. However, such high lifetimes exceed the maximum drift time in the TPC, and the corrections to the S2 signals therefore become small. 

\subsubsection{Krypton Removal System}
\label{sec::kr}

Beta decays from $^{85}$Kr or daughters of $^{222}$Rn represent a dangerous background for LXe TPCs since noble gas isotopes cannot be removed by the getter. They mix with the xenon and hence cannot be suppressed by fiducialization. Traces of krypton are removed from gaseous xenon by the same cryogenic distillation column developed and used for XENON1T~\cite{Aprile:2017aty,distillation}.  The cryogenic distillation technique leverages the higher vapor pressure of krypton than xenon at $-$98~$^\circ$C. The column can deliver natural krypton, $^{\mathrm{nat}}$Kr, over xenon  concentrations  <48\,ppq\,(mol/mol) at 90\% CL, while a single pass through the column yields a krypton reduction of $(6.4^{+1.9}_{-1.4})\cdot 10^5$. To calculate background rates we assume a $^{85}$Kr/$^{\mathrm{nat}}$Kr abundance of $2\cdot 10^{-11}$. A concentration below 360\,ppq was achieved in XENON1T by operating the distillation column in online mode, to reduce krypton while acquiring science data~\cite{XENON:2021fkt}.

All xenon gas is passed through the column once before entering the cryostat vessel. We refer to this as offline distillation. 
Conversely, the column can also be operated in an online mode, during which krypton is continuously removed from the gas volumes within the cryostat and pipes while the detector is operating. To restore equilibrium, $^{\mathrm{nat}}$Kr migrates from the liquid xenon to the now $^{\mathrm{nat}}$Kr-depleted gas phase. Section~\ref{sec::perf_rad} describes the measured performance of the krypton removal system. 

The cryogenic distillation also removes potential tritium contamination in the form of HT molecules present in the xenon. HT molecules are extremely volatile with respect to xenon. Taking measurements of helium removal in XENON1T as a benchmark for the HT case, a lower limit on the reduction factor of 144~(90\%\,C.L.) can be estimated for a single pass through the distillation column~\cite{murra}.

\subsubsection{Radon Removal Plant}
\label{sec::rn}

In XENON1T, radon emanating from materials in direct contact with the xenon was the dominant source of ER-like background in the fiducial volume through $^{214}$Pb beta decays~\cite{Aprile:2018dbl}. Radon emanation measurements performed on all XENONnT sub-systems before filling of the cryostat indicated an expected radon concentration of (4.2$^{+0.5}_{-0.7}$)\unitRad{}~\cite{XENON:2021mrg}. This initial radon concentration is about a factor~3 lower than in the XENON1T science campaigns ($\sim$13\,\unitRad{})~\cite{Aprile:2018dbl}. About 30\% of the reduction comes from the replacement of the QDrive gas pumps~\cite{qdrive}, used in XENON1T for recirculation, with new magnetically-coupled piston pumps~\cite{Schulte:2021zxc} featuring a much lower emanation rate. The remaining is a result of the extensive material selection and cleaning campaigns and of the improved surface-to-volume ratio. In addition, an online radon removal plant has been developed to further reduce this background, maximizing the XENONnT physics reach. 

As demonstrated in early tests with XENON100~\cite{RnRemovalXENON100} as well as in another dedicated setup~\cite{Bruenner:2016ziq}, cryogenic distillation can be used to effectively separate radon from xenon due to its lower vapor pressure. This technology was further explored in low-background conditions at the end of XENON1T operation by temporarily operating its krypton distillation column in reverse direction and returning the radon-depleted gas from the top of the column to the cryostat. Although only a small fraction of the recirculation flow was distilled, a $^{222}$Rn reduction of about 20\% could be achieved~\cite{XENON:2020fbs}. To fully exploit 
this concept a high-flow radon removal plant was integrated into the LXe recirculation loop of XENONnT. The plant is capable of distilling xenon extracted from both the liquid phase and the gaseous phase independently. It is described in detail in Ref.~\cite{Murra:2022mlr}.

\begin{figure}[t]
\centering 
\includegraphics[width=.48\textwidth]{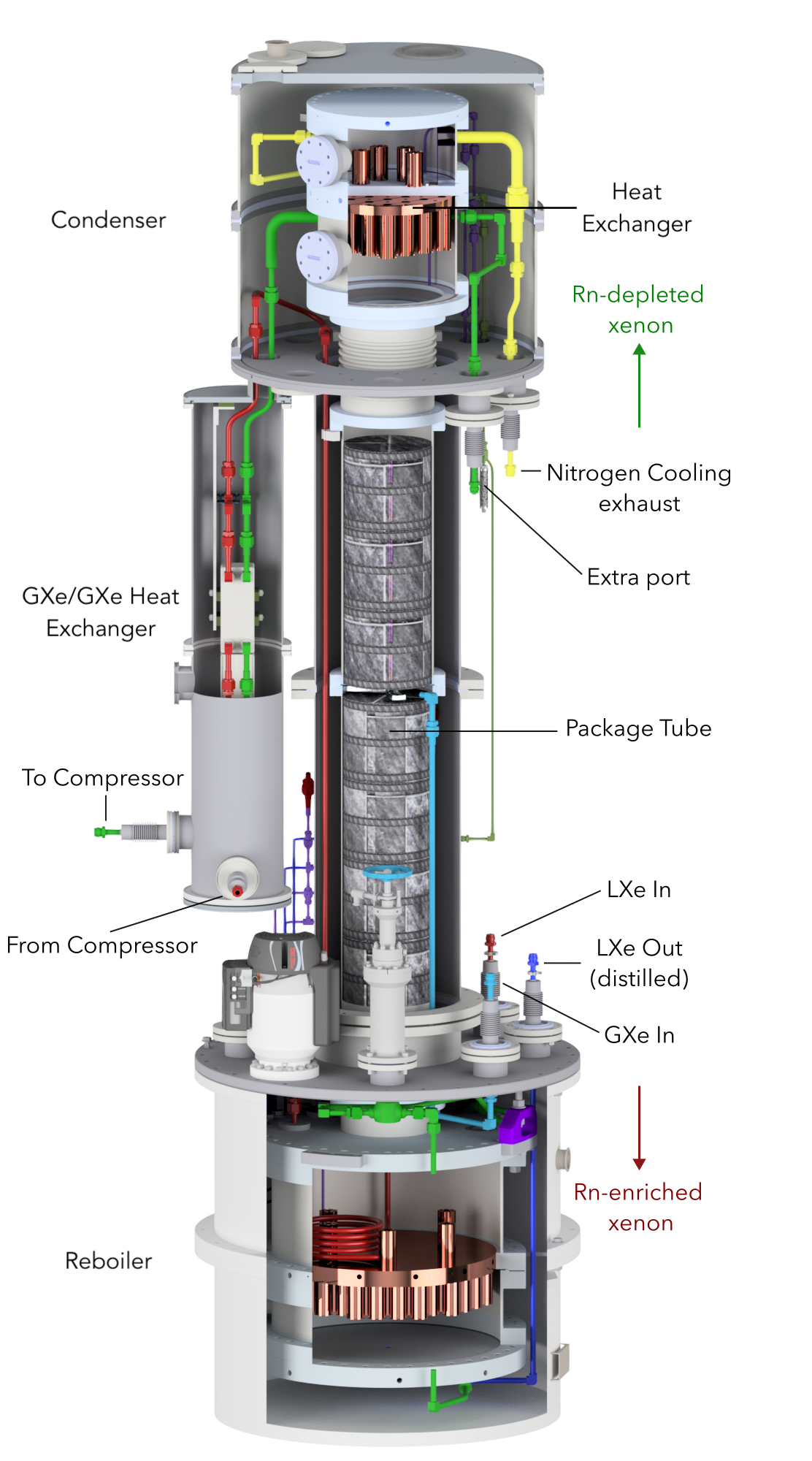}
\hfill
\caption{\label{fig:radonplant} 3D CAD rendering of the radon distillation column. Its total height is 3.8\,m and it is continuously operated to reduce the $^{222}$Rn level to the design goal.}
\end{figure}

Large distillation flows are required to efficiently extract the LXe from within the cryostat and trap $^{222}$Rn atoms into the radon removal plant before they decay inside the TPC. Given the $^{222}$Rn half-life of 3.8\,days and the full xenon target in XENONnT, a xenon flow of about 200\,SLPM (1.7\,tonnes/day) is needed to achieve a radon reduction factor of~2.1 within the active volume. The radon removal plant (Fig.~\ref{fig:radonplant}) consists of four main components: the reboiler at the bottom, the central package tube, the top condenser, and four magnetically-coupled piston pumps (not shown)~\cite{Schulte:2021zxc}. Liquid xenon from the cryostat is fed into the distillation column through the LXe purification system (see Fig.~\ref{fig:cad_cryo}). The xenon (and the radon) moves towards the bottom of the column. Once it reaches the top face of the reboiler (which also serves as heat exchanger~\cite{Murra:2022gxp}) the LXe evaporates and moves toward the top of the column, while the radon atoms due to their lower volatility accumulate in the reboiler until they decay. In order to reach the condenser located at the top of the column, the gaseous xenon passes through the package tube, providing a large surface for liquid-gas exchanges that further reduce the radon content. Once at the top condenser, which is equipped with a  LN$_2$-cooled heat exchanger, the radon-depleted xenon is partially liquefied to provide a reflux into the system, maintaining the distillation process. About 95\% of the extracted radon-depleted gaseous xenon, warmed up by a heat exchanger, is pushed by the magnetically-coupled piston pumps towards the bottom of the reboiler. This leads to a pressure increase such that the gas liquefies on the bottom surface of the reboiler, which is cooled down by the lower pressure liquid xenon evaporating on its top surface~\cite{Murra:2022mlr}. The re-liquefied xenon is then sent back to the cryostat. The remaining 5\% of the radon-depleted gaseous xenon at the outlet of the column is sent to the diving bell on top of the TPC and towards the PTR of the cryogenic system to be recondensed. The column is expected to provide a xenon/radon reduction of about~100 at the top of the column and a radon enrichment factor of about~1000 at the bottom.

The gaseous xenon above the TPC, in the warmer pipes, does not have a large mass, however it is exposed to an increased radon emanation rate from the large amount of surfaces (cables, etc) as well as from welds in the SS pipe. Therefore, the column is designed to also distill the xenon in the gas phase. This gas is directly fed into the column through a separate branch containing a QDrive gas pump~\cite{qdrive}, to drive a typical flow of 20\,SLPM, and a gas purifier~\cite{PS3-MT15-R}. According to the distillation model presented in Ref.~\cite{Murra:2022mlr} the column is close to 100\% efficient in suppressing the radon emanated into the gas phase, which contributes about 2\,\unitRad{} to the overall radon concentration~\cite{XENON:2021mrg}. The usage of this branch in combination with the high-flow distillation of the LXe allowed to reach the target concentration of 1\,\unitRad{}. Section~\ref{sec::perf_rad} describes the measured performance of the radon removal plant.

\begin{figure*}[t]
\centering 
\includegraphics[width=.95\textwidth]{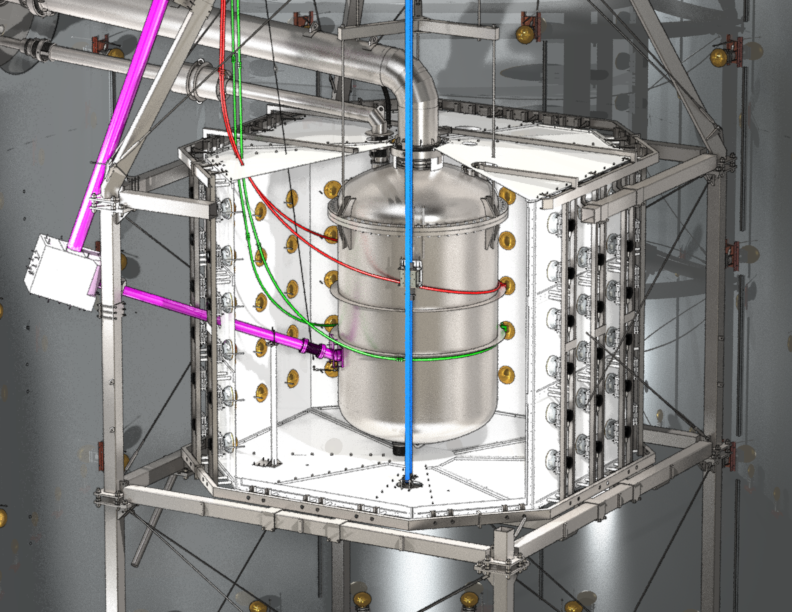}
\hfill
\caption{\label{fig:calibration}CAD rendering of the neutron veto structure (white reflectors, 120~PMTs) and the calibration system. The I-belt (blue) is used to move a tungsten collimator containing a source close to the cryostat. The L-shaped beam pipe (pink) for neutron calibrations consists of the neutron guide and beam arms, pointing towards the water tank top and the cryostat, respectively. Unshielded calibration sources can be inserted into the two U-tubes (red, green) surrounding the cryostat.}
\end{figure*}

\subsection{Veto Systems}
\label{sec::vetoes}

The central cryostat containing the TPC is surrounded by two nested active water Cherenkov vetoes that are contained in the same water tank and designed to suppress the neutron-induced NR backgrounds (see Fig.~\ref{fig:calibration}). The outer detector,  already operated in XENON1T, tags muons capable of producing high-energy neutrons that could reach the TPC and produce a NR. The inner detector, newly installed in XENONnT, encloses the cryostat and is designed to tag radiogenic or cosmogenic neutrons after their interaction inside the TPC. The system was operated during the first science run (SR0) with demineralized water only. The first doping with gadolinium to increase the neutron tagging efficiency was performed at the end of 2023.

\begin{figure}[t]
\centering 
\includegraphics[width=.49\textwidth]{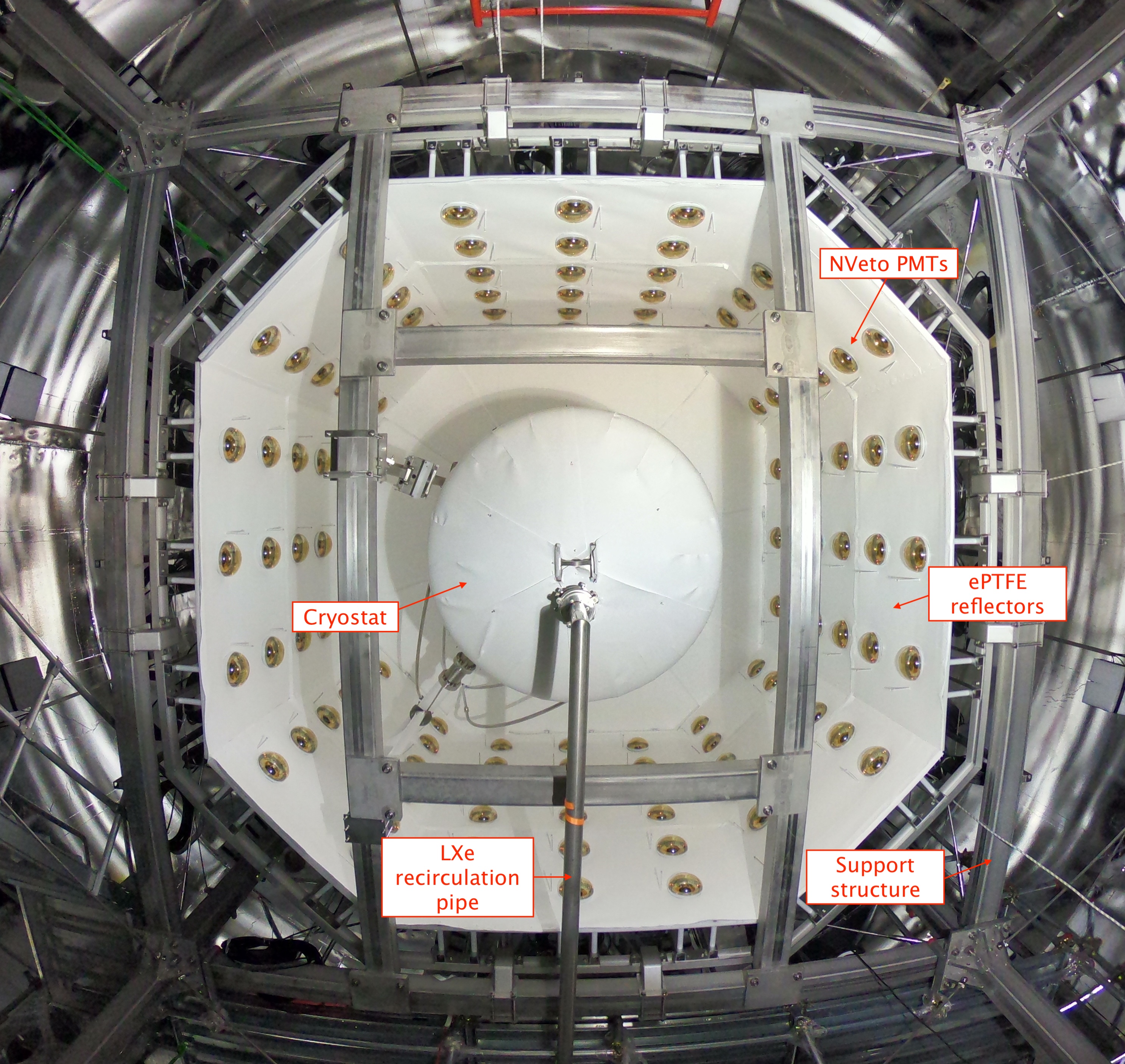}
\hfill
\caption{\label{fig:NV_photo} A picture of the neutron veto taken from below during its assembly (the bottom octagon of light reflectors was not yet in place). In the center the cryostat, covered with ePTFE sheets, and the LXe recirculation pipe are visible.}
\end{figure}

\subsubsection{Water Shield and Active Muon Veto}
\label{sec::muonveto}

The cryostat is submerged in 700~tonnes of demineralized water, kept inside a cylindrical tank of  10.2\,m height and 9.6\,m diameter. The water acts both as a passive shield for external $\gamma$-rays and neutrons, and as an active veto against cosmic muons and their induced cascades, potentially containing neutrons~\cite{Aprile:2017aty,XENON1T:2014eqx}. The water tank is turned into a Cherenkov detector by instrumenting it with 84 PMTs of 8'' diameter~\cite{R5912ASSY}, featuring $\sim$30\% QE over the wavelength range 300–600 nm. The tank surfaces are cladded with a highly reflective foil~\cite{DF2000MA} in the wavelength range of 380-1000\,nm \cite{Geis_2017}. The PMTs are immersed in water and distributed along the tank's perimeter, arranged in five vertically equidistant rings. For gain calibration purposes each PMT can be illuminated with blue LED light via a network of optical fibers. Gadolinium doping, required to operate the neutron veto that shares the same water, is expected to improve neutron tagging also in the muon veto, otherwise no detrimental effect is expected.

\subsubsection{Active Neutron Veto}
\label{sec::nveto}

The neutron veto, shown in Figs.~\ref{fig:calibration} and~\ref{fig:NV_photo}, efficiently detects neutrons interacting in Gd-doped water, as Gd is the element with the highest capture cross-section for thermal neutrons. The demineralized water of the 700~t tank will be eventually doped with 0.48\% (by mass) of gadolinium sulfate octahydrate (Gd$_2$(SO$_4$)$_3\cdot$8H$_2$O). Capture on $^{157}$Gd and $^{155}$Gd, the isotopes with the highest neutron capture cross-section, leads to the emission of $\gamma$-cascades, involving three to four photons with total energies of 7.9\,MeV and 8.5\,MeV~\cite{Beacom:2003nk,MARTI2020163549,Abe_2022}. These photons undergo Compton scattering, which frees electrons with sufficient energy to generate Cherenkov light that is detected by the PMTs. 

The neutron veto detector encompasses a volume of about 33~m$^3$ surrounding the LXe cryostat in the water shield. It consists of a lightweight stainless steel structure supporting two octagonal caps and eight side reflector panels made of high-reflectivity expanded PTFE (ePTFE) to increase light collection. These ePTFE sheets optically separate the neutron veto from the muon veto. They are located at about 1~m distance from the cryostat and are anchored to the cryostat support structure. This design was developed to accommodate all the pre-existing pipes needed for xenon handling and for the calibration tools, and to ensure sufficient water exchange with the muon veto for purification purposes. Cherenkov light emitted in the neutron veto volume is observed by 120 high-QE, low-radioactivity PMTs of 8'' diameter \cite{R5912-100-10} immersed in water. On average, they are operated at a gain of about 7$\times$10$^6$ and a dark count rate of $\mathcal{O}({1\,\mathrm{kHz}})$ above a threshold of $\sim0.3$~photoelectrons. Again, a network of optical fibers reaching each PMT allows for calibration and performance monitoring. Monte Carlo studies indicate that the mean path of Cherenkov photons from the generation point to a PMT is around 13\,m. The ePTFE sheets thus greatly increase the effective single photon detection efficiency.

A Gd-water purification system was constructed and coupled to the water tank to maintain high water transparency for Cherenkov light in the presence of dissolved Gd-sulfate. The key technology used for neutron detection was successfully developed by the EGADS group in the Super-Kamiokande collaboration~\cite{MARTI2020163549}, and replaces liquid scintillator technologies that are no longer allowed at LNGS. According to Monte Carlo simulations, based on EGADS studies and on results from XENONnT R\&D activities, such veto is expected to reach a neutron tagging efficiency of about 85\% \cite{XENON:2020kmp}. 

A total of 3.4\,tonnes of Gd-sulfate has been procured, selecting the best batch by radiopurity assays~\cite{Hosokawa:2022iuq}. It can be injected into the system via the Gd-water purification system and the water shield is not required to be emptied. Given the mild-corrosive properties of the Gd-sulfate, all materials in potential contact with the Gd-loaded water were pre-selected and tested for compatibility. During the first science runs, the neutron veto was operated with demineralized water only, to verify the performances of the system before inserting the Gd-salt. Details of the neutron veto performance during this phase are presented in Sec.~\ref{sec::commissioning}. 

\subsection{Calibration System}
\label{sec::calibration}

The calibration system is used to characterize the response of the TPC and outer detectors with a wide range of radiation sources. Primarily, three radioactive isotopes are injected, in gaseous form, directly into the GXe purification system to generate well-defined electronic recoil (ER) signals throughout the active volume and characterize the TPC response. 

The first one is $^{83\mathrm{m}}$Kr whose atoms are injected into the TPC active volume by redirecting a fraction of the GXe purification system flow through a vacuum-tight cartridge containing zeolite beads with the mother isotope $^{83}$Rb placed in between particulate filters~\cite{Manalaysay:2009yq,VHannen_2011}. These krypton atoms decay with a complex scheme leading to observable monoenergetic lines at 32.2\,keV and 9.4\,keV throughout the TPC. Since the intermediate state leading to the 9.4\,keV line has a half-life of 154\,ns, a third monoenergetic line at 41.6\,keV is observed when the decays are not resolved in time and merged in one signal. These narrow lines are ideal to correct for spatial non-uniformities of the TPC response. The rather short half-life of 1.83\,h makes this source also ideal to be used regularly during the science run, typically once every two to three weeks, to monitor the stability of the light and charge yields as well as the electron lifetime.

The second radioactive isotope is $^{220}$Rn, whose decay chain contains $\alpha$-, $\beta$- and $\gamma$-emissions. An electro-deposited $^{228}$Th source that emanates $^{220}$Rn atoms is exposed to a GXe flow used to introduce them into the TPC~\cite{Lang:2016zde,Jorg:2023nvl}. Of particular interest is the decay of $^{212}$Pb, with a beta spectrum with a $Q$-value at 560 keV to characterize the low-energy ER signal distribution in the S2 vs.~S1 parameter space. Once the $^{220}$Rn injection is terminated, the decay rate within the TPC decreases to the background level within few hours. 

The third internal radioactive source is $^{37}$Ar that decays through electron capture, releasing Auger electrons and X-rays that lead to monoenergetic lines at 2.82~keV, 0.27~keV and 0.01~keV~\cite{XENON:2022ivg} within the detector. The two lowest-energy peaks produce only a small ionization signal, which is of particular use for calibrating the charge-only low-threshold analyses~\cite{PhysRevLett.123.251801}. The highest energy peak allows to determine the detection efficiency near threshold~\cite{XENON:2022ivg}. Given the long half-life of $t_{1/2}$=35.0\,days the calibration is performed at the end of a science run, after which $^{37}$Ar atoms are removed through distillation with the krypton distillation column. A $^{37}$Ar reduction rate of about a factor~10 every 4~days was observed in XENON1T down to the original background level~\cite{XENON:2022ivg}. 

\begin{figure*}[t]
\centering 
\includegraphics[width=.95\textwidth]{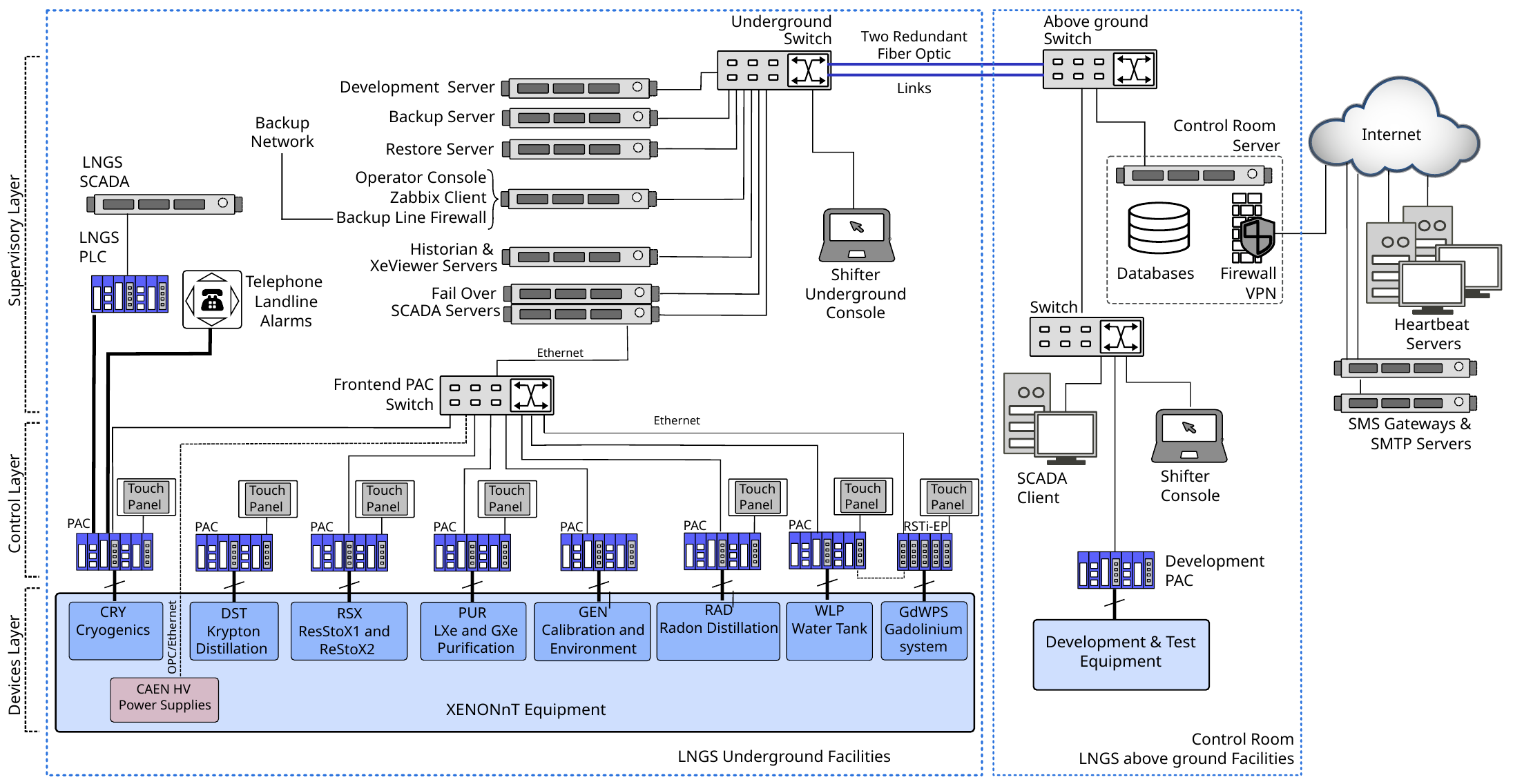}
\hfill
\caption{\label{fig:sc}Overview of the XENONnT slow control system: the information from the various sub-systems underground is acquired by Programmable Automation Controllers (PACs) and eventually stored in a database. The remote accessibility of the system is crucial for safe detector operation and continuously monitored by heartbeat servers.}
\end{figure*}

The experiment is also equipped with mechanical systems shown in Fig.~\ref{fig:calibration}, penetrating the outer detectors, to allow for the movement of solid calibration sources to within the neutron veto and close to the cryostat. The ``I-belt'' (highlighted in blue in Fig.~\ref{fig:calibration}) is used to lower a tungsten collimator which can hold a $^{88}$Y-Be photo-neutron source for low-energy NR calibration. The collimated source can be placed at various $z$-coordinates. Two independent ``U-tubes'' run around the cryostat at different heights to characterize the TPC response at various positions. They are made of 1''~diameter stainless steel tubes in which small encapsulated calibration sources are inserted with flexible guides. This system was used to calibrate the TPC and neutron veto with high-energy gamma sources ($^{228}$Th or $^{137}$Cs) and a $^{241}$Am-Be neutron source. 

An L-shaped beam pipe (pink in Fig.~\ref{fig:calibration}) can provide a collimated beam of neutrons from a neutron generator, reaching the TPC at a 20$^\circ$ angle from the \emph{xy}-plane. The pipe consists of an almost vertically aligned \emph{guide arm} of 15.3\,cm diameter, and a beam arm of 10.2\,cm diameter pointing towards the TPC.  The guide arm is used to lower a pulsed neutron generator that emits 2.5~MeV neutrons in short time pulses down to the junction point, with a maximum rate of 10$^7$\,n/s. The beam arm displaces the water in the line-of-sight toward the cryostat and acts as an effective neutron collimator. A 20\,cm thick block of borated polyethylene surrounds the junction point and captures the neutrons emitted in directions other than the beam arm. To characterize the detector response to low-energy neutrons, a small metal container to be filled with heavy water (D$_2$O) from the outside is installed in the junction point and the neutron generator can be placed next to it.  Simulations predict that high-energy 2.5\,MeV neutrons emitted by the generator and back-scattered on the D$_2$O target will reach the TPC through the beam arm with a reduced energy $\lesssim$100\,keV.

\subsection{Slow Control System}
\label{sec::slowcontrol}

The slow control  system continuously monitors the state of all sub-systems' instruments and equipment. Its software and hardware architecture is identical to the successful XENON1T system~\cite{Cardoso:2016jto,Aprile:2017aty}. It relies on industry-standard control hardware (PACs: programmable automation controllers) and software~\cite{scada}. The scalable network design was expanded to include the new systems, namely the liquid xenon purification system, radon removal plant, xenon storage (ReStoX2),  the neutron veto with its gadolinium water purification system, PMT high-voltage supplies and calibration systems. An overview of the system is shown in Fig.~\ref{fig:sc}.

\begin{figure*}[t]
\centering 
\includegraphics[width=.95\textwidth]{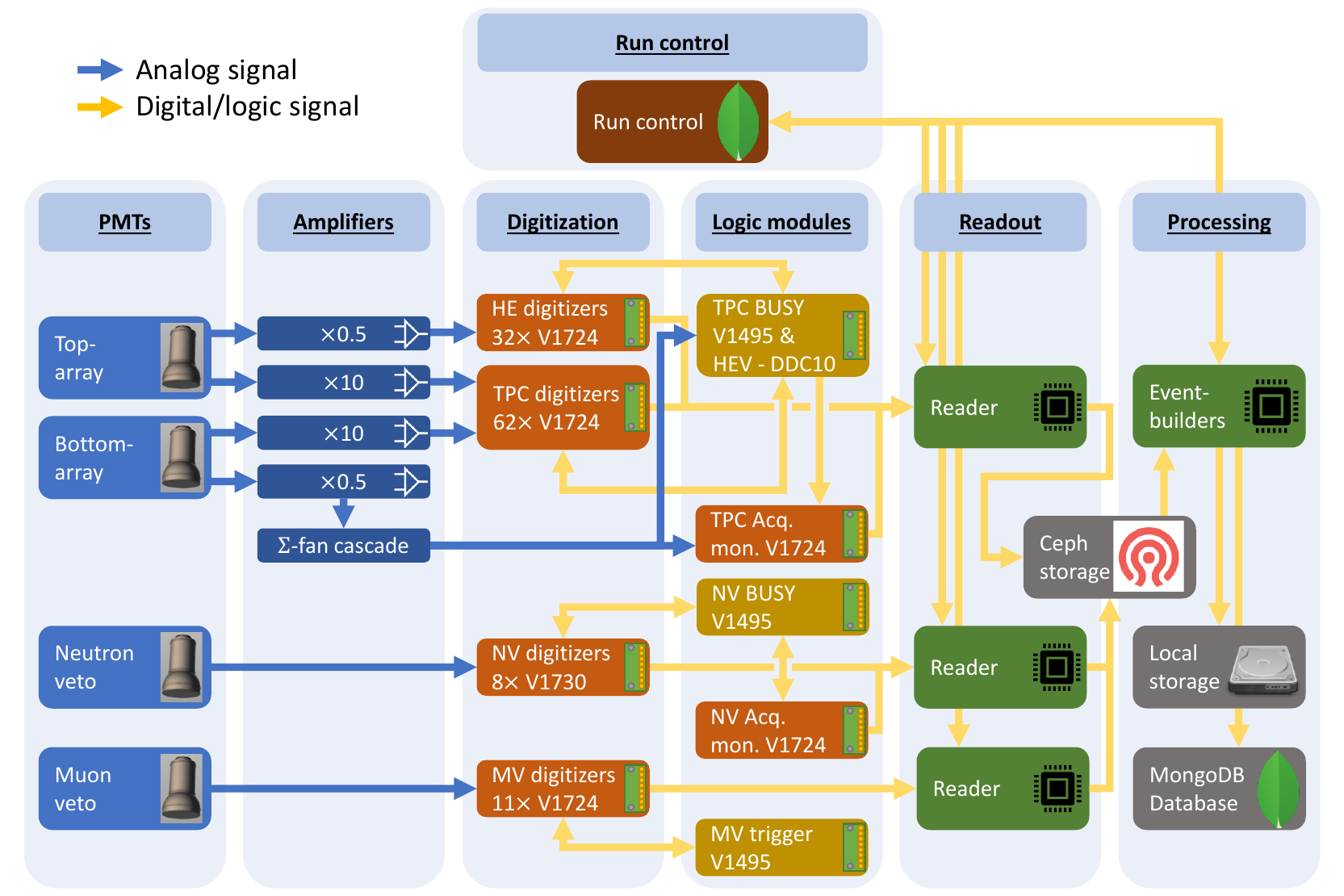}
\hfill
\caption{\label{fig:daq}Overview of the XENONnT data acquisition system~\cite{XENON:2022vye}. The PMT signals from the neutron and muon veto systems and the amplified TPC PMT signals (top and bottom array) are digitized and time-stamped before being read out. The attenuated signals (0.5$\times$) of the bottom array PMTs are summed using a cascade of linear fan-in/fan-out modules. Logic modules inhibit data readout in case of high-energy events or if digitizers are busy.}
\end{figure*}

With experience gained from XENON1T, the slow control system was upgraded to control and automate more advanced operations. Changes in the settings of the cryogenics, adjustments to PMT high voltage settings, and PMT calibration sequences can be performed using fully automated scripts. Cryogenic operation sequences control a complex network of valves and pumps based on desired running conditions. The automated calibration procedure for the 494~TPC PMTs implements predefined sequences for different calibration purposes (PMT gains, noise level, afterpulse rates) that simultaneously control the light driver, PMT high-voltage settings and the data acquisition system. A newly implemented interface to the PMT database~\cite{mongodb}, which stores voltage and current settings, allows for loading predefined voltage setting maps.

A total of about 5000~parameters are regularly monitored and stored in a Historian database~\cite{historian}. The data is used to define alarm conditions (e.g., parameter out of range, equipment failure, etc.) in which case users are notified via various communication channels. 

\subsection{Data acquisition, storage and processing}

XENONnT consists of three sub-detectors (LXe TPC, neutron veto, muon veto) equipped with a total of 698\,PMTs. The sub-detectors can be operated jointly as one system or independently. The readout (see Sec.~\ref{sec::daq}) is organized such that the data from all sub-detectors arrives in the same Ceph-based storage system~\cite{ceph} and is processed by the same data processor Strax (see Sec.~\ref{sec::event_reconstruction})~\cite{jelle_aalbers_2021_5576034,evan_shockley_2021_5591557}. This relies on a powerful computing system (see Sec.~\ref{sec::computing}).

\subsubsection{Data Acquisition System} 
\label{sec::daq}

All sub-detectors are read out using the custom-developed program Redax~\cite{XENON:2022vye}, adapted to the different requirements. With a few exceptions, XENONnT mostly relies on commercially available electronic modules and components. Fig.~\ref{fig:daq} shows an overview of the system. More details are presented in~\cite{XENON:2022vye}.

\paragraph{TPC:} All 494~analog PMT signals are fed into custom-developed dual-gain amplifiers featuring a high-gain ($\times$10) and a low-gain ($\times$0.5) output. All high-gain channels are digitized by CAEN V1724 flash ADC boards (digitizers) which provide eight channels of 100\,MHz sampling rate, 40\,MHz input bandwidth and a 2.25\,V input range with 14\,bit resolution~\cite{CAEN1724}. The DPP-DAW firmware of these modules was jointly developed with CAEN for XENON1T to enable triggerless readout~\cite{XENON:2019bth}. This means that any signal on any channel is digitized independently from all the other channels if it exceeds a pre-defined threshold (typically 15\,ADC counts corresponding to 2.06\,mV). 
Every waveform is time-stamped to allow for reconstruction of the full event; the waveform length varies depending on the length of the input signal window and includes some baseline before and after the signal. The baseline outside of these signal windows is discarded. During PMT calibration campaigns with a low-intensity external blue LED, the digitizers are operated in triggered mode, with the trigger signal being issued by the light source driver.

The low-gain outputs of the top array PMTs are also digitized individually using V1724 modules. These data avoid saturation of the digitizers by high-energy signals and can thus improve the position reconstruction for, e.g., neutrinoless double beta decay searches~\cite{XENON:2020iwh}. The low-gain outputs of the bottom array PMTs are summed up using a cascade of analog fan-in/fan-out modules. The sum signal serves as input for the high-energy veto (HEV) that can be activated during calibration campaigns to avoid digitizing and storing high energy events. The sum signal is also additionally digitized by a V1724 with a high threshold, that ensures that it will never be in a busy state. This unit is referred to as the acquisition monitor. It is not vetoed and thus allows for the TPC activity to be monitored even when data recording is inhibited on all other digitizers due to a global busy signal.

All TPC and neutron veto digitizers share a common 50\,MHz clock signal distributed by a module which also delivers GPS-time stamped signals~\cite{DeDeo:2019cih}. A crate controller~\cite{CAENV2718} generates a logical start signal that is distributed simultaneously over all the digitizers, thereby allowing for absolute timestamps on every channel. The digital busy signals (LVDS) provided by the digitizers in case their internal buffer is full are fed into a logic module~\cite{CAENV1495} which vetoes further acquisition on all channels once one digitizer sends a busy signal. At the start and the end of the busy periods, digital signals are sent to the acquisition monitor to allow for a precise calculation of the dead time.

The HEV is realized by a 100\,MHz digitizer~\cite{skutek} with an FPGA and an on-board Linux system. It continuously digitizes a copy of the low-gain sum signal from the bottom array PMTs and identifies large S2~signals using a rolling window filter in real-time that extracts signal parameters. 
If the signal parameters exceed pre-defined thresholds, a digital veto signal is issued to the logic module~\cite{CAENV1495}, which inhibits data acquisition during the veto signal. The length of the veto signal can be set to depend on the signal parameters or to a fixed length; the start and end of the veto signal are also recorded by the acquisition monitor.  

To achieve a high-data throughput above 0.5\,GB/s, the TPC digitizers and the acquisition monitor are read out in parallel by three ``reader'' servers~\cite{Fujitsu-2CPUs-192}. Two additional servers~\cite{Fujitsu-2CPU-64} serve as backup readers. Redax reads data from digitizer boards, transforms the data into the Strax data processor format (see Sec.~\ref{sec::event_reconstruction}), compresses it, and eventually pushes it to a Ceph network storage disk~\cite{ceph} made of twelve write-intensive SSDs with 900\,GB capacity each. The entire triggerless readout system was designed such that essentially every signal in the TPC during dark matter runs, appearing in any PMT at any time, is read out and stored without any pre-selection. This gives maximum flexibility for data processing and facilitates the lowest possible thresholds. The raw data stream is not discarded, i.e., the original information can be retrieved at a later point and re-analyzed. 

\paragraph{Neutron Veto:} The signals from the 120~neutron veto PMTs are digitized without additional amplification by 8~CAEN~V1730 digitizers~\cite{CAEN1730}. These 16\,channel modules provide 500\,MHz sampling rate, 250\,MHz bandwidth and 2.0\,V input range at 14\,bit resolution. The faster digitization frequency compared to the TPC was chosen to exploit the fast time structure of the Cherenkov signals. The digitizers use the same trigger-less firmware as the ones in the TPC. Time synchronization with the rest of the detector is realized as described above. The neutron veto busy is again issued by a logic module~\cite{CAENV1495}; the busy information is stored by the neutron veto acquisition monitor consisting of a V1724 digitizer. 

A reader server~\cite{Fujitsu-1CPU-128} continuously pulls the data from the neutron veto digitizers and its acquisition monitor. The Redax program is again used here; it handles the data in the same way as described for the TPC. To reduce the amount of data, a software trigger is implemented in Strax (see Sec.~\ref{sec::event_reconstruction}): it searches for a pre-defined number of PMT signals occurring in coincidence within a certain time window and discards all other data from the reconstruction process.

\paragraph{Muon Veto:} The readout of the 84~muon veto PMTs is mostly unchanged compared to XENON1T~\cite{XENON:2019bth}: the unamplified signals are digitized by CAEN V1724 digitizers operated in a global trigger mode. A trigger is issued by a logical unit once $N_\mathrm{pmt}$\,signals occur within a certain time window~\cite{CAENV1495}. Time synchronization with the TPC and the muon/neutron veto is achieved by storing the trigger signal on the vetoes' acquisition monitor (that is running on the synchronized 50\,MHz clock). Upon triggering, the waveforms from all PMTs in a 5.12\,\mus{} window around the trigger are stored. The muon veto digitizers are also read using Redax running on one reader server~\cite{Fujitsu-1CPU-8}.

\paragraph{Data Acquisition Control:} The data acquisition systems of the three sub-detectors are controlled via a single web-based platform with user access control. Data acquisition can be started in ``linked mode'', i.e., synchronized across all systems or individually for, e.g., calibration purposes. Standard acquisition settings can be selected as pre-defined use cases. Information on the runs, which are identified via a unique run number, are stored in a run database, built with MongoDB~\cite{mongodb}. The platform also provides live information on the status of the various components, including the TPC, and allows for browsing the run history.

\subsubsection{Event Reconstruction}
\label{sec::event_reconstruction}

The goal of the event reconstruction is to determine the properties of physical interactions in the detector from the waveforms recorded by the individual PMTs. The untriggered raw data stream from the data acquisition system consists of temporal sequences (one for each PMT) of fragments of digitized waveforms. It is processed in real-time by custom-developed Python-based packages, Strax and Straxen, used to analyze waveforms from all three sub-detectors~\cite{jelle_aalbers_2021_5576034,evan_shockley_2021_5591557} (for simplicity both packages will be later referred as Strax). Although Strax would allow for live data reduction and hence for the discarding of initial raw data files, a copy of the original DAQ data stream is stored on a local disk at LNGS to allow for successive offline reprocessing (see Sec.~\ref{sec::computing}).  

The event reconstruction is performed by Strax in multiple steps. \emph{Hits}, corresponding to significant excursions with respect to the baseline, likely to be associated with real photoelectron signals or electronic noise spikes, are identified within each time sequence. Starting from hits, defined channel by channel, \emph{peaklets} are subsequently identified by requiring a coincidence of at least $N_\mathrm{hits}$ hits among at least $N_\mathrm{pmt}$ photomultipliers within a specific time window~$T$. While the parameters may be chosen differently for specific analyses, typical values for regular S1+S2 analyses are: $N_\mathrm{hits}$=3, $N_\mathrm{pmts}$=3, and $T$=50\,ns. Each peaklet is then classified as an S1~signal \emph{peak} or part of an S2 signal. The algorithm also tags unidentified peaklets which are likely induced by noise. Due to the large longitudinal diffusion caused by the long drift lengths, S2~signals often consist of multiple S2~peaklets. S2~peaklets and neighboring signals are thus merged into a single S2~\emph{peak}. Exploiting the causal time sequence of S1 and S2~peaks and the maximum drift time, \emph{events} are identified. At this stage, the definition of events is purposely kept rather broad and also includes interactions with only a lone S1 (S2) signal or S1 and S2 signals in more complex combinations, e.g., from multiple interactions. 

Variables associated with each event (\texttt{event\_info}) are then used to further refine the type of interactions of interest, required for the specific analysis (standard S1+S2 analysis, charge-only analysis, etc.). The advantage of the triggerless solution adopted by XENON is the possibility to heavily modify, even at much later stages, the entire peak identification logic (e.g., $N_\mathrm{hits}$, $N_\mathrm{pmt}$, $T$ and other criteria) without information losses which inevitably occur in hardware trigger scenarios.

\subsubsection{Computing model and infrastructure}
\label{sec::computing}

At the XENONnT site, online processing is performed for data quality monitoring and immediate feedback (see Fig.~\ref{fig:daq}). In addition, the digitized PMT waveforms from all three detectors are stored for subsequent offline reprocessing. The latter uses customized settings for the reconstruction algorithms with more accurate performance as well as updated corrections and calibration constants. 

The large data throughput of about 0.8\,PB/year is dominated by regular high-statistics calibration campaigns. XENONnT requires a highly distributed data processing scheme, designed to minimize the time needed for massive reprocessing of full data campaigns. The processing pipeline leverages on the US~Open Science Grid (OSG) and the European Grid Infrastructure (EGI). Once raw data arrive at the local storage at LNGS they are automatically transferred to Grid-accessible remote storage elements (RSE) both in Europe (Surfsara Nikhef, CCIN2P3, INFN-CNAF) and the USA (UChicago MWT2, UCSD Expanse), with all being accessed through the LHCOne private network~\cite{lhcone}. The data is managed by a custom code, developed on the backbones of the Rucio software~\cite{rucio2019}, that interfaces with the Run Database and allows automatic replication of data at various RSEs while keeping track of their location. For long-term data storage, a copy of the raw data is written to tape at CNAF. 
The CI Connect service at the University of Chicago serves as a unique job submission portal and the usage of HTCondor~\cite{https://doi.org/10.1002/cpe.938} and glideinWMS services~\cite{glideins} allows for seamless access to all available computing resources, both on the Grid and at local clusters~\cite{osti_10064984}. Chunks of raw data, together with Singularity Containers~\cite{Container} embedding the software environment and the Strax code to be executed, are sent to remote nodes that return a series of outputs files after processing (see Sec.~\ref{sec::event_reconstruction}). These outputs, once returned to the primary submission node, are merged and tagged with a unique run-based identifier. The Pegasus~\cite{deelman-fgcs-2015} workflow manager streamlines the pipeline and automatically resubmits interrupted jobs ensuring that all processing jobs are properly completed. 

High-level data files are made available to the entire XENON collaboration on a dedicated cluster that supports Jupyter notebook analyses. Since Strax provides different output data types, which correspond to different processing stages (see Sec.~\ref{sec::event_reconstruction}), the developed pipeline accommodates reprocessing starting from intermediate stages as well. This allows for significantly accelerated reprocessing (if handling raw data can be avoided). Monte Carlo data production and the associated analysis follow a similar pipeline.

\section{Detector Performance}
\label{sec::commissioning}

All XENONnT detectors and sub-systems went through a period of commissioning that lasted several months. In this section we report on the detector performance in the configuration of the experiment used during SR0, searching for new physics in low-energy electronic recoils~\cite{PhysRevLett.129.161805} and for WIMP dark matter~\cite{XENON:2023cxc}.

\subsection{Performance of the Vetoes}
\label{sec::vetoperf}

In the commissioning phase and SR0 the water tank, and hence the neutron and muon vetoes, were operated with demineralized water, to check the overall performance of the system before doping with Gd-salt. 

\begin{figure*}[t!]
\centering 
\includegraphics[width=.82\textwidth]{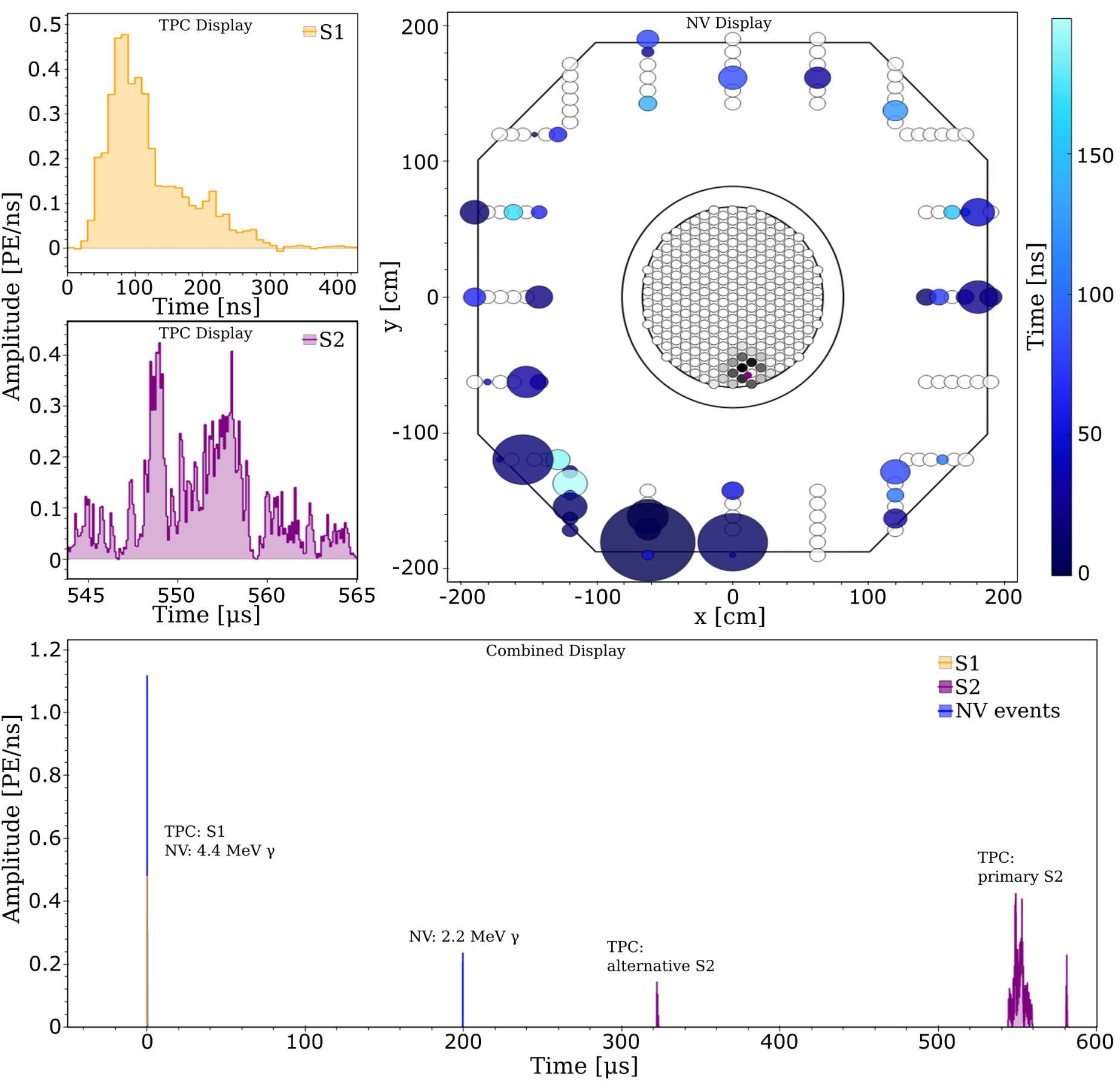}
\hfill
\caption{\label{fig:combined_event_display} Signals detected during a calibration with an Am-Be source. They are associated to a neutron interaction in the TPC and a coincident 4.4\,MeV $\gamma$-photon interacting in the neutron veto (filled with demineralized water), followed by a 2.2\,MeV $\gamma$ from subsequent neutron capture in the water. The bottom combined display summarizes the time correlation between these events in the two detectors: at $t=0$ a neutron is emitted; the associated $\gamma$-photon interacts through Compton scattering in the water, inducing Cherenkov light, which is detected by neutron veto PMTs within few tens of~ns. At the same time, the neutron scatters in the xenon target producing a nuclear recoil and inducing the emission of prompt light (S1). The back-scattered neutron leaves the TPC and, upon reaching the NV, is moderated and captured by a proton, inducing the emission of a 2.2~MeV photon detected by the neutron veto PMTs after $\sim$200\,\mus. Within the TPC two S2-type signals are detected. Properties of the S1 and of the S2-like signals are used to define the primary S2 that, in this case, is the one recorded around 550\,\mus{}. The alternative S2 is compatible with a single electron signal. The S1 and S2 signals are shown with higher time resolution in the top left two panels. 
The top-right plot shows a top view of TPC and neutron veto, with the hit pattern for the S2 signal as seen by the TPC top PMT array (see central circular panel; the magenta dot indicates the reconstructed S2 position) as well as the hit pattern from the detection of the 4.4\,MeV $\gamma$-photon in the neutron veto PMTs. The staggered circles indicate the PMTs of the neutron veto from the bottom (inner) to the top (outer circles). The circle size is proportional to the detected signal and the color code shows the time of detection.}
\end{figure*}

\subsubsection{Muon Veto Performance}
\label{sec::MVperf}

Despite the presence of the neutron veto at its center, the muon veto shows detection performance similar to that measured during XENON1T operations~\cite{Aprile:2017aty}. This is possible thanks to the usage of high-reflectivity ePTFE panels installed on the outer surface of the neutron veto. The PMTs are operated at an average gain of 6$\times$10$^6$ to allow for efficient detection of single photoelectrons above the noise level. The detector was operated with a hardware trigger requiring at least 5~PMTs detecting at least 1\,PE each within a 300\,ns coincidence window. MC simulations show that the detection efficiency for crossing muons is $\sim$100\% with this trigger, while the recorded event rate of $\sim$10\,Hz leads to a negligible dead time. According to the simulations, the same configuration also yields a 50\% tagging efficiency for muon-induced neutrons from an electromagnetic or hadronic shower generated by a muon interacting outside of the water tank. When used as veto, the muon veto detector induces a small loss around 1\% to the TPC live time during regular science runs.

\subsubsection{Neutron Veto Performance}
\label{sec::NVperf}
 
The key feature of the neutron veto is to measure events in coincidence with a low-energy nuclear recoil in the TPC, thereby allowing for the identification and suppression of events caused by neutron backgrounds. An example of such an event is shown in Fig.~\ref{fig:combined_event_display}, where a neutron from an Am-Be neutron source undergoes a single scatter within the TPC, leaves the cryostat and is then moderated and captured in the neutron veto. The emission of the neutron from the Am-Be source is accompanied by a 4.4~MeV $\gamma$-ray which overlaps in time with the nuclear recoil S1~signal observed inside the TPC. The back-scattered neutron is eventually captured by a proton under the emission of a 2.2~MeV $\gamma$-ray. Finally, the S2 signal from the initial NR in the TPC is observed. 

\begin{figure}[t]
\centering 
\includegraphics[width=.48\textwidth]{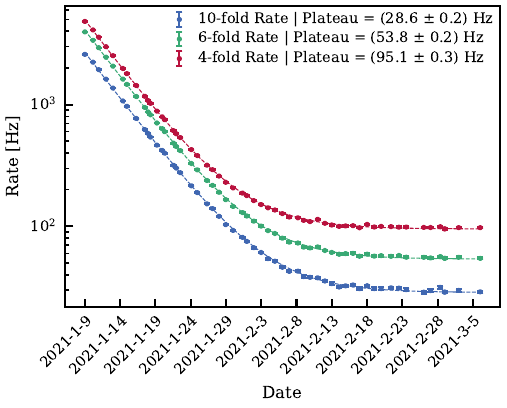}
\caption{\label{fig:nveto_coincedence_rate} Event rate in the neutron veto as a function of time for 4~(red), 6~(green), and 10-fold~(blue) PMT coincidence after filling the water shield with demineralized water. The initial high rate due to contamination with natural $^{222}$Rn decreases consistent with its half-life, after which the rates stabilize to below 100~Hz even for the 4-fold coincidence requirement.}
\end{figure}

The neutron veto event rate when requiring a 4-, 6-, and 10-fold PMT coincidence logic is shown in Fig.~\ref{fig:nveto_coincedence_rate}. The initial coincidence rates are well above 1\,kHz, which is caused by the presence of $^{222}$Rn in the water used to fill the water tank. The rates decrease thereafter with a half-life consistent with the $^{222}$Rn half-life. After the $^{222}$Rn decayed, the coincidence rates reached a plateau at 100\,Hz or lower.

The neutron tagging efficiency is measured to be ($53 \pm 3$)\%, when requiring a coincidence time window of 250\,\mus{} between the S1 signal in the TPC and the Cherenkov signal in the neutron veto \cite{XENON:2023cxc}. Thanks to the low counting rate in the neutron veto, the corresponding live time loss in the TPC, induced by random coincidences, is limited to 1.6\%. Once the water is Gd-doped, the neutron capture time will be significantly reduced, while about 8~MeV is emitted in $\gamma$-rays. 

\subsection{Performance of Cryogenics and Purification Systems} 
\label{sec::purperf}

This section summarizes the steps taken to fill the cryostat with LXe and describes how the xenon was treated in order to achieve low levels of electronegative and radioactive ($^{85}$Kr, $^{222}$Rn) impurities, and to avoid possible contamination with $^3$H, which could affect the low-energy electronic recoils background~\cite{PhysRevD.102.072004,PandaX:2022ood}.

\subsubsection{Processing of the Xenon Gas}

The xenon inventory was treated to remove krypton and possible tritium contamination as much as possible before filling the detector. The 3.2~tonnes of xenon previously used in XENON1T already had a low krypton concentration and were stored in the newly built ReStoX2  storage system. The additional 5.7~tonnes of xenon, required to fill the XENONnT cryostat and to operate its purification systems, were processed through the krypton~distillation column and also stored in ReStoX2. Before distillation each new bottle of xenon was screened with an~RGA plus cold-trap system to avoid introducing high levels of impurities into the clean storage. Once all xenon was stored in ReStoX2, the ReStoX1 storage vessel containing only GXe was treated to reduce any internal H$_2$O contamination to minimal levels: the gas was heated to $\sim$40$^{\circ}$C and was continuously purified using the gaseous purification system for about two weeks. This procedure also reduced the potential contamination from tritiated water (HTO). Subsequently, the total xenon inventory was transferred from ReStoX2 to ReStoX1, via the purifiers of the gaseous purification system to reduce potential $^3$H contaminations. In fact, all xenon gas entering XENONnT was passing through a hot gas purifier at least twice. 

\subsubsection{Cryostat Filling}

Several preparatory operations were conducted to reduce the outgassing of O$_2$ and other gases of the plastic components of the TPC (mostly PTFE) before filling the cryostat with xenon. After evacuating the cryostat with turbomolecular pumps for about four weeks, it was filled with GN$_2$ and partially evacuated in multiple cycles. During this period of 5~weeks, in-situ measurements of the radon emanation rate of the entire cryostat and TPC system were performed. After the radon emanation rate measurements were completed, the system was evacuated with turbomolecular pumps for a second period of four weeks. Finally, the cryostat was filled with GXe which, while being at room temperature, was continuously purified using the gaseous purification system for approximately two weeks.

The cryostat and the TPC were then slowly cooled down to LXe temperature over the course of two weeks to maintain limited temperature gradients across fragile components, such as the TPC electrodes. The first 2.2~tonnes of LXe were filled by condensing xenon gas from ReStoX1 into the cryostat with the PTR and the LN$_2$ systems, at a rate of about 200~kg/day, until the bottom PMT array and the cathode electrode were submerged in LXe. The remaining 6.3~tonnes of LXe were filled using LXe from ReStoX1, going through its heat exchanger to vaporize the LXe and recondense the clean GXe after its passage through the gas purifiers to remove possible $^3$H contamination. In this way the cryostat could be filled at a rate of approximately 500~kg/day. This filling mode was used only after the fragile components of the TPC were submerged in LXe to avoid thermal shocks, since the pipes returning LXe to the TPC are located directly above the windows of the bottom array PMTs. 

Since this initial filling, the system has been operated at a nominal pressure of 1.9~bara by maintaining the copper block connected to the PTR at $-$98$^{\circ}$\,C via a heating element. In this way, the pressure inside the cryostat is regulated to a precision of about 10\,mbar. A control loop taking one of the capacitive LXe height measurements as a process value adjusts the GXe flow to the TPC bell to maintain the LXe level at the desired height between the gate and anode electrodes. The LXe level fluctuations achievable with this control system are less than 20\mum, causing no measurable effects in the S2-signals.

\subsubsection{Performance of the Purification Plant} 
\label{sec::perf_pur}

\begin{figure}[t]
\centering 
\includegraphics[width=.48\textwidth]{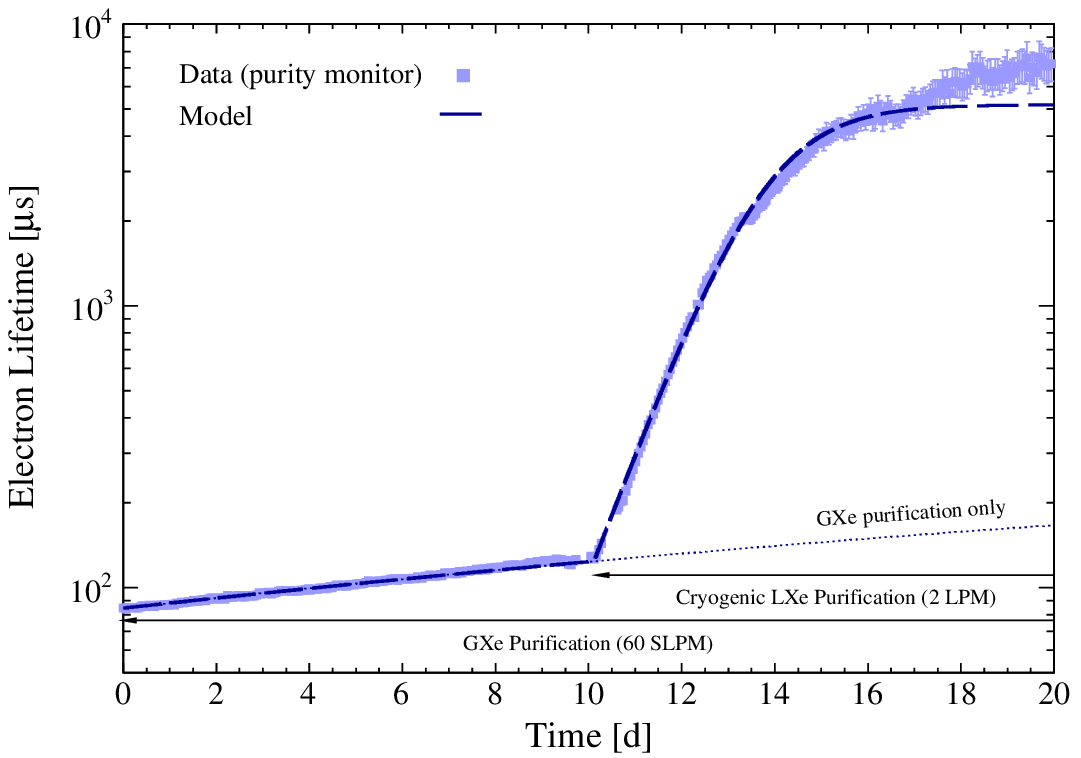}
\caption{\label{fig:e_lifetime}Electron lifetime evolution as measured with the purity monitor. A clear improvement occurs after the start of cryogenic LXe purification operated at 2~LPM with the high-efficiency O$_2$ purifier (Q-5). The model assumes a purifier efficiency of 100\% and an O$_2$ outgassing rate of 0.11\,mg/day.}
\end{figure}

During the filling procedure outlined above, the total xenon inventory was passed through the gas purification system a second time to minimize the initial concentration of electronegative impurities, such as H$_2$O, dissolved in the LXe. A few days after the filling operation was completed the liquid xenon flow through the LXe purification system was initiated, with the purifier units bypassed. This allowed for measuring the concentration of electronegative impurities via the purity monitor. An initial electron lifetime $\tau_e$=71\,\mus{} was measured without purification. This is significantly higher than the $\tau_e\sim3$\,\mus{} measured shortly after filling XENON1T, as a result of the more thorough preparation of the TPC's plastic components. Purification of the LXe and GXe volumes in the cryostat was subsequently started with the gaseous purification system, with flows of about 58~SLPM and 2~SLPM, respectively. LXe was vaporized, and the purified GXe was recondensed by heat exchangers as described in section~\ref{sec::xenon}. Figure~\ref{fig:e_lifetime} shows the evolution of the electron lifetime as measured by the purity monitor, which samples LXe from the cryostat, upstream of the LXe purifier unit. After operating the gaseous purification system for two weeks $\tau_e\sim100$\,\mus\, could be achieved. 

The steep rise in Fig.~\ref{fig:e_lifetime} shows the improvement achieved when the LXe flow was directed to the LXe purifier unit. The LXe was directly extracted from the bottom of the cryostat and sent to a custom-designed purifier  operated at cryogenic temperatures without being evaporated.  In this initial phase, the Q-5~purifier (see Sec.~\ref{sec::LXe_pur}) was used to quickly improve the purity. A typical flow of 2~LPM achieved a   $\tau_e\sim6$\,ms within a few days. In preparation for the low-background science campaign, the Q-5~purifier was substituted by a filter filled with St707 SAES pills which feature a reduced radon content and was used throughout SR0. Operation with this new purifier made it possible to reach electron lifetimes greater than 10\,ms, greatly exceeding the TPC's maximum drift time of about 2\,ms.

\subsubsection{Performance of the Krypton and Radon Removal Systems}
\label{sec::perf_rad}

The initial krypton content in the LXe after filling was about $^{nat}$Kr/Xe=480~ppq, as measured by rare gas mass spectroscopy (RGMS)~\cite{rgms}. This concentration was higher than desired, especially when compared to the targeted background contribution from $^{214}$Pb. It was further reduced to a concentration of $^{nat}$Kr/Xe = (56$\pm$ 36)\,ppq, well below the design level of 100\,ppq, by operating the krypton distillation column in ``online'' mode for six weeks~\cite{PhysRevLett.129.161805}. The ability of XENONnT to run the krypton distillation column while operating the detector is unique and guarantees a low krypton concentration over extended periods of operation.

At this point, the measured $^{222}$Rn concentration in the TPC, without the radon distillation system, was about 3.5\,\unitRad{}, which is slightly below the expected concentration based on the material screening results (see Sec.~\ref{sec::materials}). The effective material screening and cleaning campaigns, together with an expected reduction of the radon concentration due to the improved surface-to-volume ratio in XENONnT, contributed to a lower radon concentration compared to XENON1T. During SR0, the radon distillation column was operated in ``gas-only'' mode, i.e., without purifying xenon taken out of the liquid, resulting in a further reduction of the radon concentration to (1.81\,$\pm$\,0.02)\,\unitRad{}. The radon level in the ongoing science run (SR1) is even below 1\,\unitRad{} with the radon column operated in combined liquid-and-gas mode.

\subsection{TPC Performance}
\label{sec::TPCperf}

The commissioning of the TPC started after the cryostat was filled with xenon and passed through two phases, with and without water shielding. In the initial phase without water shielding only basic TPC functionality was tested. Otherwise, the high rate of interactions inside the TPC, especially in combination with operations affecting electron drift and signal multiplication, would have led to high light levels potentially damaging the PMTs. The commissioning in this phase was thus performed only with a few active PMTs when the electrodes were under high-voltage; the electrodes were independently tested up to intermediate voltages. The cathode grid was operated at $-$15\,kV while keeping the gate grid at ground, reaching an average drift field of $\sim$100\,V/cm, slightly exceeding the drift field of XENON1T~\cite{Aprile:2018dbl}. Similarly, in the absence of a drift field, a potential difference of 4.5\,kV was applied across the anode-gate region to confirm the production of electroluminescence. Higher voltages were not tested given the impossibility of detecting potential micro-discharges efficiently.

Once the cryostat was immersed in water, the TPC was characterized in low-background conditions as reported in the next sections. In this phase, during a cathode ramping up at around $-$12\,kV, a discharge happened and the bottom screening electrode and cathode were found to be in electrical contact then. Without opening the cryostat the most likely explanation is that the electrical contact is caused by one or more electrode wires that broke during the incident. This prevented stable operation of the cathode beyond $-$2.75\,kV, limiting the maximum drift field to $\sim$23\,V/cm corresponding to a maximum drift time of 2.3\,ms. While larger electric fields are preferred for minimizing electron capture from electronegative impurities, the high electron lifetime $\tau_e > 10$\,ms achieved in XENONnT enables operations also at low fields. 
First results from SR0~\cite{PhysRevLett.129.161805} indicate that the negative impact of low fields can be successfully mitigated, for ER searches, at the level of the analysis. 
In the search for NR WIMP signals the lower drift field has an impact on the S2/S1-based discrimination and the ability to distinguish single from multiple interactions in the target. Nevertheless,  XENONnT could obtain competitive WIMP results in SR0~\cite{XENON:2023cxc}.

In low-background conditions, the voltage across the anode and gate electrodes was limited to 4.6\,kV. At higher voltages, intense single-electron (SE) emission, with rates four orders of magnitude higher than for lower voltages, was observed close to one of the anode transverse wires.  The extra peaks in the data slowed down data acquisition and real-time processing and the prolonged high light levels could be dangerous for the PMTs. Similar anomalous emissions were observed in the past in other xenon setups and are usually referred to as ``hot spots''; their origin is likely multifaceted but still not fully understood.~\cite{LUX:2020vbj,Tomas:2018pny}. 
At lower voltages, a milder version of the phenomenon (``warm spot'') did occasionally appear which neither impacted the XENONnT DAQ performance nor data analysis (see Sec.~\ref{sec::S2respperf}). Occasionally, it transitioned into a hot spot which required the electrodes to be powered off for some time and consequently led to a live time loss of $\sim$11\%.

In SR0, the TPC electrodes were biased with $V_\mathrm{c}$=$-$2.75\,kV, $V_\mathrm{TFSE}$=$+$0.65\,kV, $V_\mathrm{g}$=$+$0.3\,kV and  $V_\mathrm{a}$=$+$4.9\,kV, establishing electric drift and extraction fields of $E_\mathrm{d}$=23.0$^{+0.4}_{-0.3}$\,V/cm and $E_\mathrm{e}$ between 2.9\,kV/cm and 3.7\,kV/cm in the liquid (depending on the radial position due to the sagging of the electrode plane) in the 4-tonnes fiducial volume defined in~\cite{XENON:2020kmp}. The bottom and top screening electrodes were biased with V$_\mathrm{bS}$=$-$2.75\,kV (due to the electrical connection with the cathode) and V$_\mathrm{tS}$=$-$0.9\,kV. 

Figure~\ref{fig:combined_event_display} shows a typical PMT waveform produced by a low-energy nuclear recoil in the TPC during calibration with an Am-Be neutron source. The $z$-coordinate of the interaction is derived from the drift time~$t_\mathrm{drift}$ defined by the time difference between the S2 and S1~signals. The observed event's position in the horizontal plane is inferred from the distribution of the S2~signal across the PMTs of the top array by a neural network algorithm trained with Monte Carlo simulations. A $t_\mathrm{drift}$-dependent correction is applied to account for the divergence of the drift field and to define $X^\mathrm{rec}$ and $Y^\mathrm{rec}$, the reconstructed coordinates of the original interaction (see Sec.~\ref{sec::S2respperf}). The spatial coordinates of the interaction are used to define the variables~cS1 and~cS2, obtained by correcting~S1 and S2 with light collection efficiency (LCE) maps extracted from calibration data capturing the spatial non-uniformity of the TPC response.

\subsubsection{Photomultiplier Performance} 
\label{sec::pmtperf}

Throughout SR0 the PMTs were operated at constant voltages that were individually chosen to ensure high single photoelectron (SPE) acceptance while minimizing dark count rate and light emission. None of the PMTs were biased beyond $-$1.5\,kV to avoid instabilities such as transient flashes. The final gains ranged from 1.2$\times$10$^6$ to 2.4$\times$10$^6$, with an average of 1.87$\times$10$^6$ at a standard deviation of 0.35$\times$10$^6$. Calibrations with blue LED light, guided into the TPC through optical fibers, were performed every week and the gains were found to be stable within 1\% over SR0. The SPE acceptance is the fraction of the SPE signals above the hardware ADC digitization threshold of a given channel (typically set to 15 ADC counts, see Sec.~\ref{sec::daq}); the average SPE acceptance was (93.0$\pm$1.8)\%. A typical dark count rate of $\sim$40\,Hz was measured at LXe temperature for all PMTs. 

Seventeen PMTs out of 494 were excluded from SR0 analysis due to sub-optimal performance: 11~developed leaks at cryogenic temperature leading to frequent HV~trips, 4~showed light emission,  gain instabilities or noise and 2~could not be operated due to cabling issues. Of the remaining PMTs used for SR0 23  show signs of vacuum degradation, presenting (still minimal) enhancements of either the Xe$^{+}$- or N$_2^{+}$-induced after-pulse rates. While the increasing xenon after-pulse rates are symptoms of leaky PMTs, the nitrogen contamination is likely due to emanation from PMT materials that trapped gas during the fabrication. The vacuum degradation for these 23~PMTs varies significantly, with increases in after-pulse rates ranging from 0.01\%/month to 0.14\%/month.

\subsubsection{TPC Light Response} 
\label{sec::S1respperf}

\begin{figure}[t]
\centering 
\includegraphics[width=.48\textwidth]{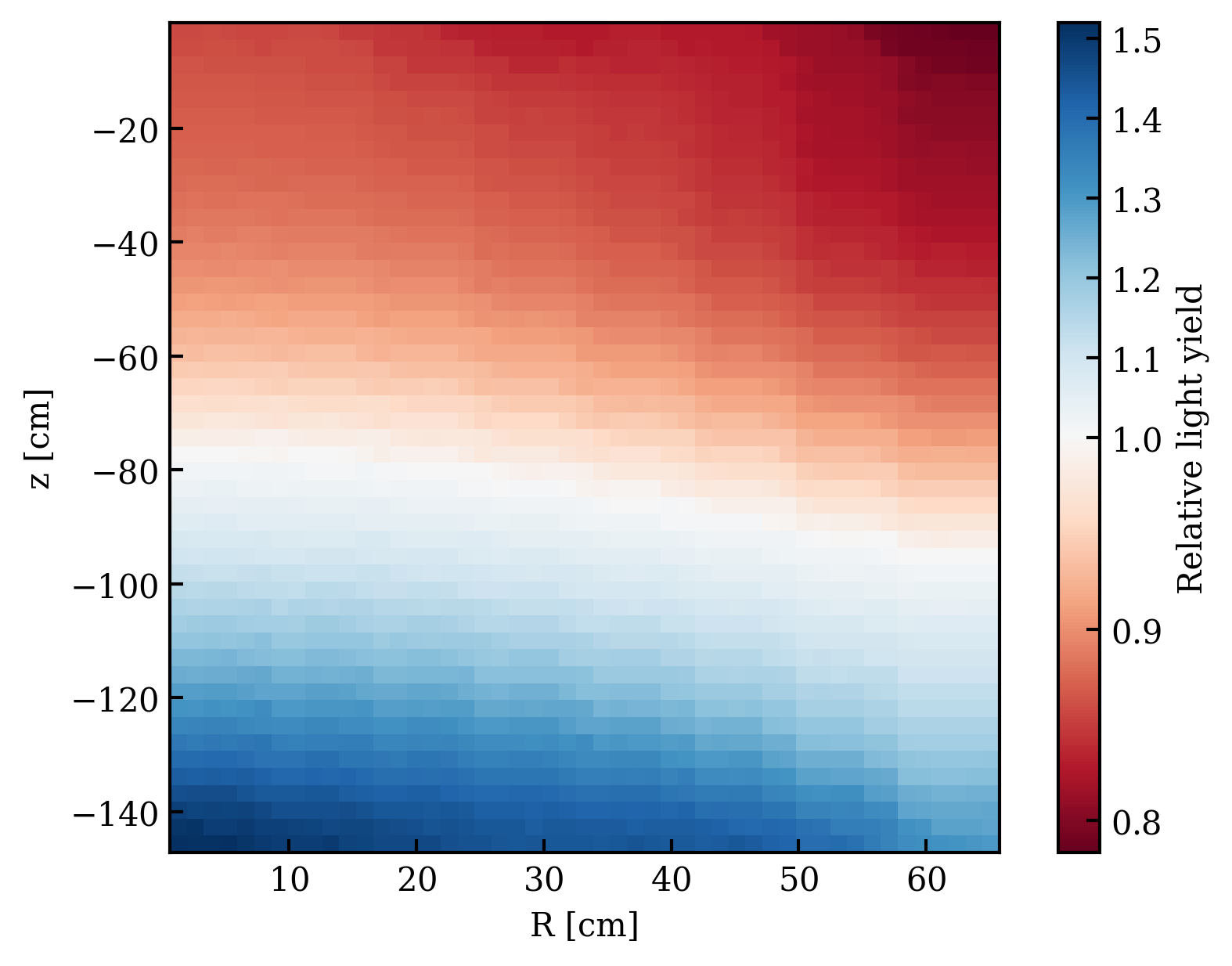}
\caption{\label{fig:83kr_spatial_s1} Light yield variation as a function of the reconstructed interaction depth and radius derived from the 41.6\,keV $^{83m}$Kr summation line. The reference frame is such that the liquid/gas interface and cathode planes are located at $z$=0~cm and $z\sim-$149~cm, respectively.}
\end{figure}

The light response of the TPC, characterized by the LCE map, was measured using the isotope $^{83m}$Kr. The isotope is mixed into the purified gaseous xenon that, once liquefied, is injected into the TPC's active volume from below the cathode. After 2 hours from the start of the operation, the $^{83m}$Kr atoms are uniformly distributed throughout the entire volume. Figure~\ref{fig:83kr_spatial_s1} shows the light yield variation derived from the 41.6\,keV sum energy line as a function of the reconstructed interaction depth $Z^\mathrm{rec}$ and radius $(R^\mathrm{rec})^2=(X^\mathrm{rec})^2+(Y^\mathrm{rec})^2$. No PMT QE correction has been applied. The observed variation is caused by a combination of solid-angle coverage and reflectivity on the TPC walls and the liquid-gas-interface. Three-dimensional S1 LCE maps for the top and bottom PMT arrays are generated in the same way and used to define the cS1~variable, corrected for spatial non-uniformity, and normalized to the average S1~value in the active volume. 
Measurements of the light yield are reported in Sec.~\ref{sec::energyrespperf}.

\subsubsection{TPC Charge Response} 
\label{sec::S2respperf}

\begin{figure}[t]
\centering 
\includegraphics[width=.48\textwidth]{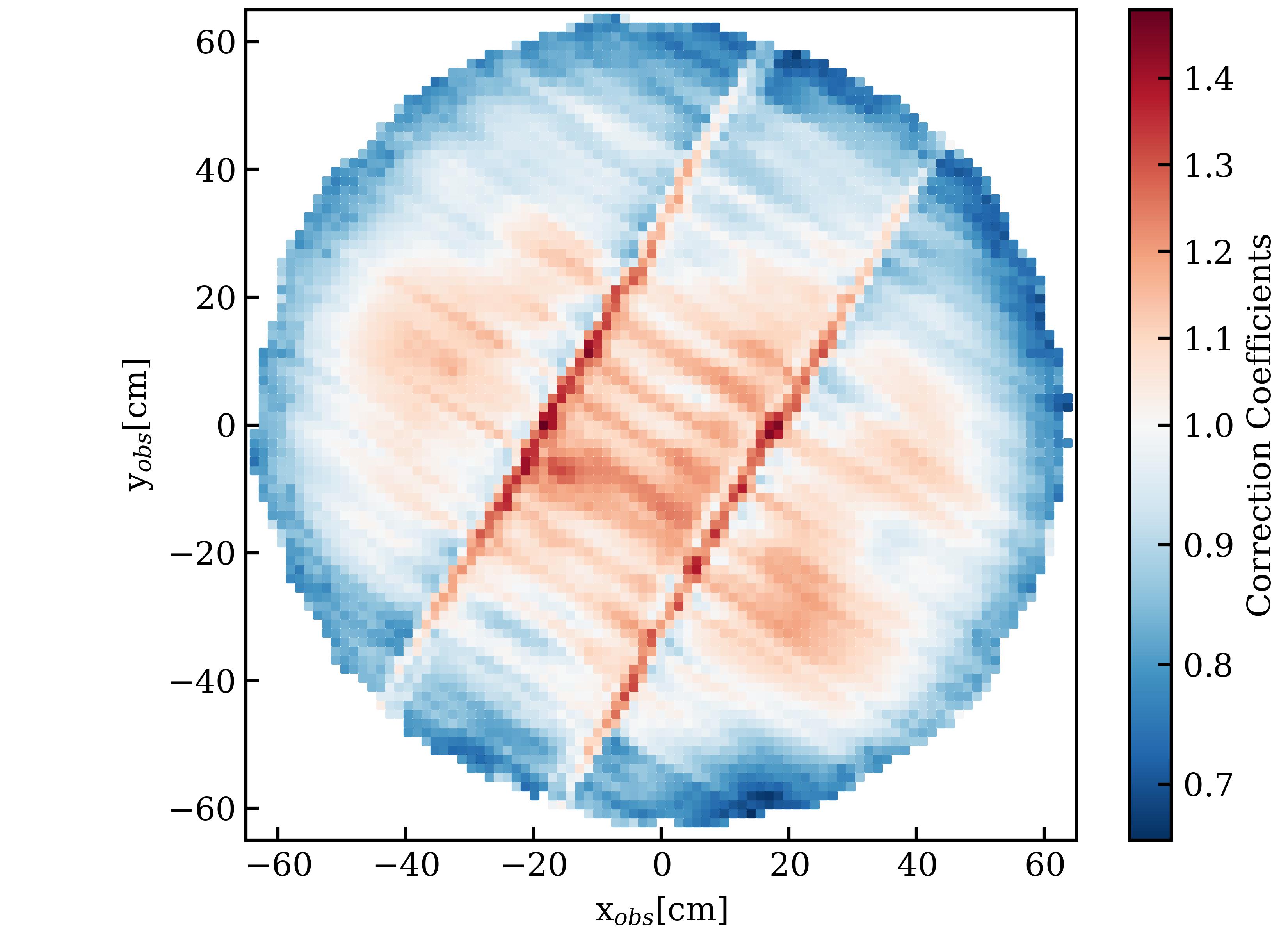}
\includegraphics[width=.48\textwidth]{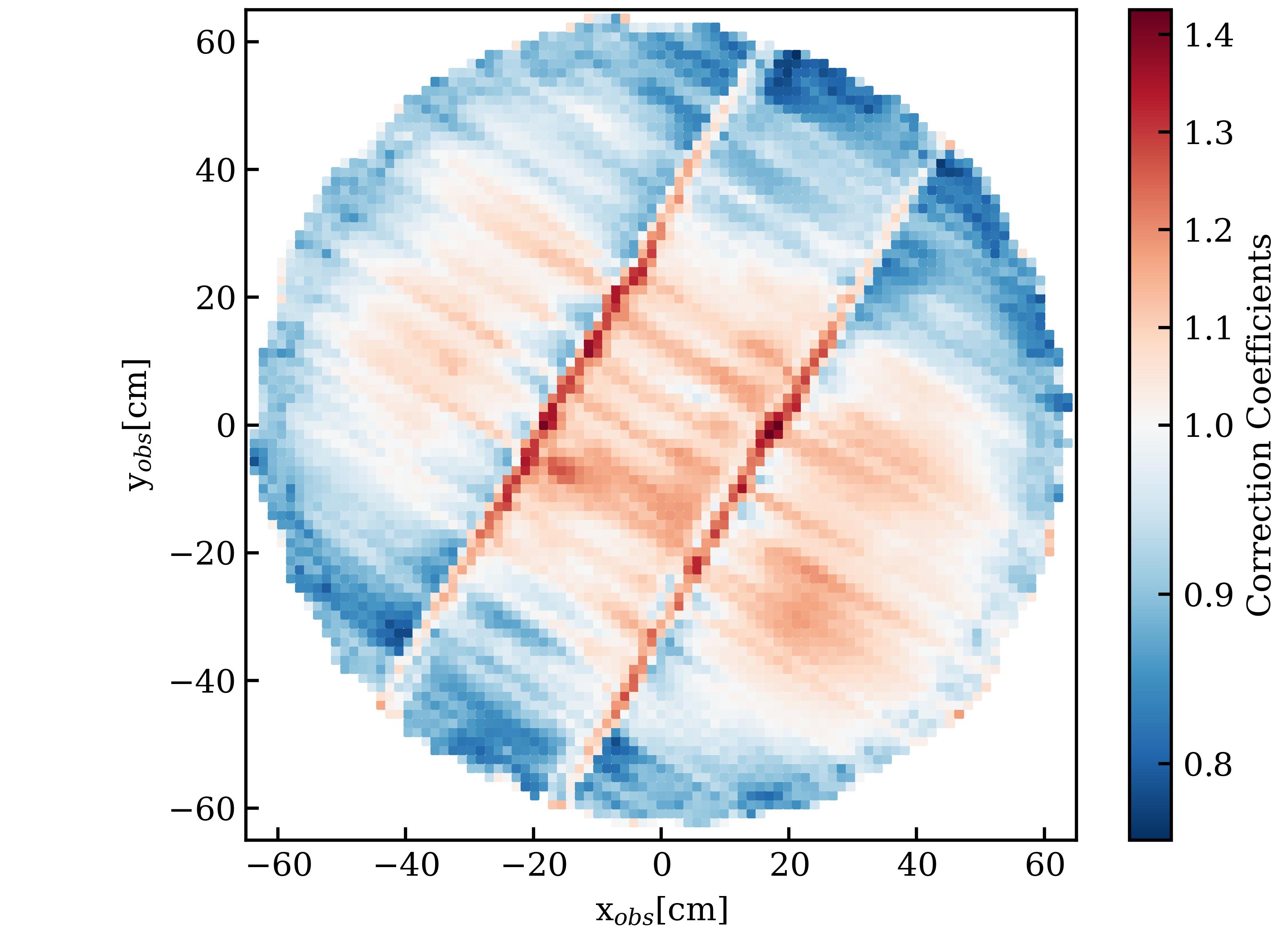}
\caption{\label{fig:83kr_spatial_s2} Relative S2 correction maps for the top array (top) and bottom array (bottom) as a function of the reconstructed coordinates $X_\mathrm{obs}$ and $Y_\mathrm{obs}$ derived from $^{83m}$Kr data. The two evident narrow bands with enhanced S2 response are produced by the transverse wires located on the gate and anode electrodes to mitigate their sagging.}
\end{figure}

The charge response of the TPC was also characterized using monoenergetic $^{83m}$Kr signals. Due to field non-uniformities the observed S2 position can differ from the interaction position. Therefore the observed (uncorrected) coordinates $X_\mathrm{obs}$ and $Y_\mathrm{obs}$ are used to produce the relative S2 correction maps for the top and bottom array shown in Fig.~\ref{fig:83kr_spatial_s2}. The two visible bands correspond to the location of the transverse wires installed on the anode and gate grid to reduce sagging. Electric field simulations indicate that the field next to these wires is greater than average, which leads to an enhanced electroluminescence signal. 
This is due to the larger electron extraction efficiency of $>$70\% around these wires compared to $\sim$53\% in the rest of the TPC.
These two-dimensional LCE maps are used to define the cS2~variable, corrected for $XY$-spatial non-uniformity, and normalized to the average value of the S2~signal. 

\begin{figure}[t]
\centering 
\includegraphics[width=.48\textwidth]{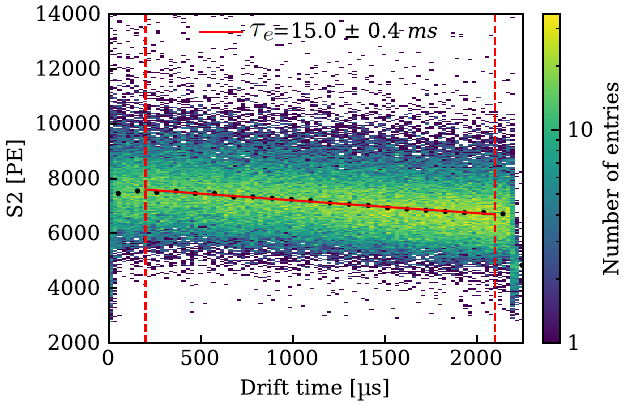}
\caption{\label{fig:83kr_z_s2} 2D histogram of the S2 area (with $XY$~spatial corrections applied) as a function of drift time for $^{83m}$Kr decays. In red, overlaid to the 2D histogram, a negative exponential fit of the average S2 area values (black points) used to extract the electron lifetime $\tau_e = (15.0\pm 0.4)$\,ms for this run.}
\end{figure}

The charge loss due to electron attachment to electronegative impurities leads to a $Z$-dependence of the charge signal. It is monitored and corrected using $^{83m}$Kr as well; Figure~\ref{fig:83kr_z_s2} shows the cS2 signal area before the correction as a function of the drift time, corresponding to the $Z^\mathrm{rec}$ coordinate. The electron lifetime extracted for this run is $\tau_e$=(15.0$\pm$0.4)\,ms which leads to a relatively small correction given the maximum drift time of about 2.3\,ms. 

The uniform distribution of the krypton atoms inside the TPC was used to quantify the 
drift field non-homogeneity and to compare it to the COMSOL simulation; details can be found in~\cite{XENONnT:2023dvq}. The observed and simulated radial distributions were matched assuming a charge distribution on the reflective PTFE panels, similar to~\cite{LUX:2017mhm}. An overall surface charge density $<$0.5\,\mmu{}C/m$^2$ was obtained. The simulated electric field matching the observed distribution has an average strength of $22.6^{+0.4}_{-0.6}$\,V/cm inside the fiducial volume defined in~\cite{XENON:2020kmp}. The sub-optimal potential on the cathode electrode causes some electric field lines, mostly located at the bottom edge of the TPC, to end at the TPC wall. This leads to a charge-insensitive volume of around 112\,kg, less than 2\% of the active target.

\subsubsection{Light and Charge Yield} 
\label{sec::energyrespperf}

\begin{figure}[t]
\centering 
\includegraphics[width=.49\textwidth]{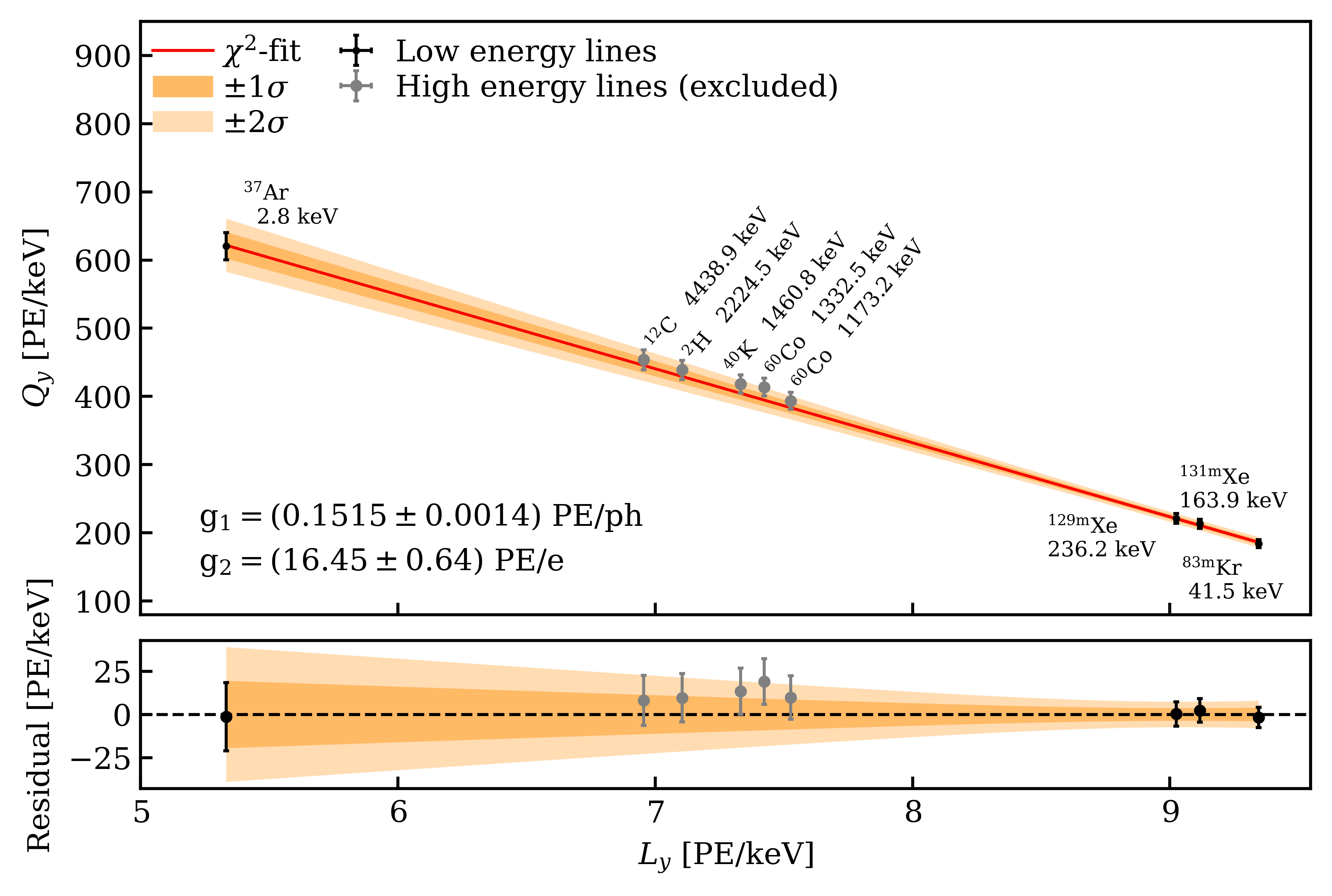}
\caption{\label{fig:doke} Light yield $L_y$ and charge yield $Q_y$ (in PE/keV) for several monoenergetic ER lines between 2.8\,keV ($^{37}$Ar~\cite{XENON:2022ivg}) and 4.4\,MeV ($^{12}$C). The values from low-energy lines $^{37}$Ar, $^{83m}$Kr, $^{131m}$Xe and $^{129m}$Xe (the latter two activated during neutron calibration) are fitted with the function $Q_y=-\frac{g_2}{g_1}L_y+\frac{g_2}{W}$ derived from Eq.~(\ref{eq::ces}). The best-fit result is shown in red. Higher-energy lines were excluded from the fit since they are more affected by potential reconstruction biases.}
\end{figure}

The detector-specific gains $g_1$ (photon detection efficiency) and~$g_2$ (charge gain) describe the number of detected photoelectrons (PE) per scintillation photon or ionization electron generated by a primary ER interaction depositing some energy~$E$:
\begin{equation}\label{eq::ces} E= W \left( \frac{\textrm{cS1}}{g_1} + \frac{\textrm{cS2}}{g_2} \right) \textrm{,}
\end{equation} 
where $W$=13.7\,eV is the average energy required to generate a scintillation photon or an ionization electron~\cite{XENON:2020iwh}. The gains $g_1$ = (0.152 $\pm$ 0.002)\,PE/$\gamma$ and $g_2$ = (16.5 $\pm$ 0.6)\,PE/$e^-$ are derived by measuring the light yield $L_y$ and charge yield $Q_y$ (in PE/keV$_\textrm{ee}$) for several monoenergetic lines between 2.8\,keV ($^{37}$Ar) and 236.2\,keV ($^{129m}$Xe), covering the region of interest for WIMP searches. $L_y$ and $Q_y$ were regularly monitored with several sources and were stable within 1.0\% and 1.9\%, respectively. Figure~\ref{fig:doke} shows these measurements and a fit with a function derived from Eq.~(\ref{eq::ces}). A bias of 1-2\% in the reconstruction of the line energies is included in the systematic uncertainty of the fit. The higher-energy lines are not used in the fit, however, the yields derived by taking them into account agree with the best-fit result.

\section{Conclusions}
\label{sec::outlook}

XENONnT is taking data since the start of the commissioning campaign in fall 2020. A first science run (SR0), with run conditions and detector performance as described in Sec.~\ref{sec::commissioning}, was conducted from July 6, 2021 to November 10, 2021. The total acquired live-time was 97.1~days. Thanks to the various background reduction measures and the successful operation of the systems presented in this article, a low-energy electronic recoil background of $(15.8 \pm 1.3)$\, events/(tonne$\times$year$\times$keV$_\textrm{ee}$) was achieved~\cite{PhysRevLett.129.161805}. This corresponds to a factor~5 reduction compared to XENON1T, and it is the lowest electronic recoil background level ever achieved in a dark matter experiment. A first dark matter search using the same dataset did not observe an excess of events above the background expectation, leading to a minimum upper limit on the WIMP-nucleon scattering cross section of $2.6\times 10^{-47}$\,cm$^2$ for a WIMP mass of 28\,GeV/c$^2$ at a 90\% confidence 
level~\cite{XENON:2023cxc}. Since then, XENONnT has continued to acquire science data with a further decreased background from $^{222}$Rn to a level below 1\,\unitRad. At the end of 2023, the water tank was doped with an initial batch of 500\,ppm of Gd-sulphate octahydrate salt, to improve the performance of the neutron veto reducing the background from radiogenic neutrons.

\begin{acknowledgements}
We would like to thank R.~Adinolfi, G.~Bucciarelli, K.~Deweese, A.~Goretti, R.~Leguijt,  M.~Linvill, C.~Orr, M.~Tobia, and R.~Walet for their contributions to the realization of the XENONnT experiment.
We gratefully acknowledge support from the National Science Foundation, Swiss National Science Foundation, German Ministry for Education and Research, Max Planck Gesellschaft, Deutsche Forschungsgemeinschaft, Helmholtz Association, Dutch Research Council (NWO), Fundacao para a Ciencia e Tecnologia, Weizmann Institute of Science, Israeli Science Foundation, Pazy Foundation, Binational Science Foundation, Région des Pays de la Loire, Knut and Alice Wallenberg Foundation, Kavli Foundation, JSPS Kakenhi and JST FOREST Program ERAN in Japan, Tsinghua University Initiative Scientific Research Program, DIM-ACAV\textsuperscript{+} R\'egion Ile-de-France, and Istituto Nazionale di Fisica Nucleare. This project has received funding/support from the European Union’s Horizon 2020 research and innovation program under the Marie Skłodowska-Curie grant agreement No 860881-HIDDeN. We gratefully acknowledge support for providing computing and data-processing resources of the Open Science Pool~\cite{OSP} and the European Grid Initiative~\cite{EGI}, in the following computing centers: the CNRS/IN2P3 (Lyon - France), the Dutch national e-infrastructure with the support of SURF Cooperative, the Nikhef Data-Processing Facility (Amsterdam - Netherlands), the INFN-CNAF (Bologna - Italy), the San Diego Supercomputer Center (San Diego - USA) and the Enrico Fermi Institute (Chicago - USA). We acknowledge the support of the Research Computing Center (RCC) at The University of Chicago for providing computing resources for data analysis.

We are grateful to Laboratori Nazionali del Gran Sasso for hosting and supporting the XENON project.
\end{acknowledgements}


\bibliographystyle{spphys2} 
\bibliography{xent_instrument}  

\end{document}